\documentclass[pdflatex,sn-mathphys-num,iicol]{sn-jnl}

\usepackage[T1]{fontenc}
\usepackage[utf8]{inputenc}
\usepackage{graphicx}
\usepackage{mathtools,amssymb,bm,mathrsfs,booktabs}

\newcommand{\rev}[1]{#1}
\newenvironment{revision}{}{}

\begin{document}

\title[Effective matter conversion in gravitational collapse]
{Effective Matter Conversion in Gravitational Collapse and the Dynamical Formation of Regular Black Holes}

\author*[1]{\fnm{Emmanuele} \sur{Battista}}
\email{ebattista@lnf.infn.it}
\email{emmanuelebattista@gmail.com}

\author[2]{\fnm{Ali} \sur{{\"O}vg{\"u}n}}
\email{ali.ovgun@emu.edu.tr}

\author[3,4]{\fnm{Vitalii} \sur{Vertogradov}}
\email{vdvertogradov@gmail.com}

\affil[1]{\orgname{Istituto Nazionale di Fisica Nucleare, Laboratori Nazionali di Frascati},
  \orgaddress{\postcode{00044}, \city{Frascati}, \country{Italy}}}

\affil[2]{\orgdiv{Physics Department}, \orgname{Eastern Mediterranean University},
  \orgaddress{\street{via Mersin 10}, \city{Famagusta}, \postcode{99628},
  \state{North Cyprus}, \country{T{\"u}rkiye}}}

\affil[3]{\orgname{Wilczek Quantum Center, Shanghai Institute for Advanced Studies},
  \orgaddress{\city{Shanghai}, \postcode{201315}, \country{China}}}

\affil[4]{\orgname{University of Science and Technology of China},
  \orgaddress{\city{Hefei}, \postcode{230026}, \country{China}}}

\abstract{\rev{We study inverse source reconstruction in generalized Vaidya
spacetimes. A prescribed density fixes the mass and tangential pressure,
while a two-sector decomposition determines a dimensionless radial balance
function. We distinguish this function from a time-directed conversion rate
and identify the additional null-flux information required for a covariant
exchange vector. Positivity restricts the allowed target pressures: a de
Sitter core cannot be represented by nonnegative sectors with nonnegative
tangential equations of state. An explicit finite-density profile admits a
positive vacuum-like completion, finite curvature invariants, and inner and
outer trapping horizons above a calculable threshold. Its total source
satisfies the null, weak, and dominant energy conditions during monotonic
accretion, while the timelike convergence condition fails in the core. We
check polytropic, bag-model-inspired, and condensate-inspired profiles and
the corresponding cosmological reconstruction. Finally, we compute the
stationary endpoint's shadow and compare its exterior deformation with
published Sagittarius A* measurements. The construction establishes local
curvature regularity and marginal-sphere formation, without claiming a
microscopic formation mechanism, perturbative stability, or geodesic
completeness.}}

\maketitle
\section{Introduction}
\label{sec:intro}
\begin{revision}
Spherical collapse is a useful setting in which to distinguish the formation
of trapped surfaces from the appearance of a curvature singularity. The
Oppenheimer--Snyder solution illustrates collapse to a black hole
\cite{Oppenheimer:1939ue}, while the singularity theorems establish geodesic
incompleteness under specified energy and global assumptions
\cite{Penrose:1964wq,Hawking:1973LSS}. Geodesic incompleteness does not, by
itself, require a divergent scalar curvature invariant.

Regular black-hole metrics replace the central curvature divergence by a
finite-density region, usually approaching de Sitter geometry
\cite{Bardeen:1968,Dymnikova:1992ux,Hayward:2005gi,Ansoldi:2008jw,Lan:2023rbh}.
Their sources include nonlinear electromagnetic fields, anisotropic stresses,
and effective vacuum contributions
\cite{Ayon-Beato:1998hmi,Bronnikov:2000yz,Bronnikov:2000vy}.
The distinction between a regular metric and a consistent dynamical source
remains essential. Limiting-curvature constructions and analyses of global
structure already showed that local regularity need not settle the causal
completion of the spacetime \cite{Frolov:1988vj,Borde:1996df}. Regular black holes
have also been studied through their shadows, rotating extensions, and
perturbative spectra \cite{Vagnozzi:2022moj,Abdujabbarov:2016hnw,Toshmatov:2017zpr,
Toshmatov:2015wga}, which provides motivation for future observational and
gravitational-wave diagnostics of dynamically generated regular interiors.  Further insight into singularity avoidance has also been obtained from
Lorentzian--Euclidean black-hole models, where the emergence of an atemporal
region can regularize the geometry and modify the causal structure of the
black-hole interior \cite{Capozziello:2024ucm}. More generally, recent analyses
of families of regular spacetimes have clarified how regularity requirements
are tied to the energy conditions, providing a useful diagnostic framework for
distinguishing physically admissible regular geometries from purely formal
metric constructions \cite{Wang:2026jvo,Wang:2026sqr}.
Recent work has examined collapse in higher-curvature theories, including
higher-dimensional settings, alongside four-dimensional time-dependent
interiors and transitions between singular and regular geometries
\cite{Bueno:2024zsx,Bueno:2024eig,Bueno:2025gjg,Ovalle:2025pue,
Borissova:2025msp,Borissova:2025hmj,Muniz:2025ugk}.

Beyond their role as nonsingular geometries, regular black holes have become important testing grounds for strong-field phenomenology. Their shadows, lensing properties, optical signatures, and limiting-curvature behavior have been investigated in rotating, charged, scale-dependent, and nonlinear-electrodynamic models \cite{Abdujabbarov:2016hnw,Toshmatov:2017zpr,He:2023bme,Balart:2024rtj,Koch:2025gaw,UktamjonUktamov:2026dep}. Complementary analyses of massive-scalar, electromagnetic, Dirac, and gravitational perturbations—including quasinormal modes, excitation factors, time-domain evolution, and late-time tails—have clarified how regular cores and surrounding matter distributions can modify black-hole spectra \cite{Balart:2023odm,Davlataliev:2024mjl,Lutfuoglu:2025mqa,Lutfuoglu:2026boa,Lutfuoglu:2026zel,Lutfuoglu:2026etg,Rahmatov:2026qow}. Further information is provided by thermodynamic and Hawking-radiation studies, as well as by periodic-orbit waveforms and the dynamics of charged or magnetized particles around regular black holes \cite{Ali:2022tdt,Ditta:2024jrv,Rayimbaev:2023vzk,Alloqulov:2025bxh}.

A de Sitter core violates the timelike convergence condition, but can preserve
the null convergence condition. The latter is the condition used in the
Penrose trapped-surface theorem. Consequently, violation of the strong energy
condition alone does not establish complete collapse without singularities.
The global extension, the existence of a noncompact Cauchy surface, and the
behavior of an inner horizon must also be examined
\cite{Penrose:1964wq,Borde:1996df,Borissova:2025msp,Borissova:2025hmj}.
We keep these global questions separate from the local construction below.

Dense-matter equations of state provide useful tests of candidate sources.
Quark matter and condensate profiles motivate particular pressure or density
relations \cite{Witten:1984rs,Farhi:1984qu,Chavanis:2011uv,Harko:2011zt},
but neither motivation supplies a covariant dynamical completion automatically.
Earlier effective collapse models introduced matter-sector redistribution
\cite{Vertogradov:2025dust,Vertogradov:2025collapse,Vertogradov:2025jxp,
Vertogradov:2025snh,Vertogradov:2024seh}.
In particular, finite-density QCD closures can still give singular cores, and
positive baryon--radiation mixtures cannot reproduce vacuum-like central
pressure \cite{Lambiase:2026hza}. Thus a successful algebraic reconstruction
must be distinguished from a positive-density material interpretation.

We study this distinction in the generalized Vaidya class
\cite{Husain:1995,WangWu1999,Vertogradov:2016gc}. Its stress tensor contains an anisotropic
sector with radial pressure $P_r=-\rho$ and an independent null flux. Given a
target density, the Einstein equations determine the mass and tangential
pressure. A two-sector decomposition then determines a dimensionless radial
function $\beta(v,r)$. This is an inverse construction: the evolution is
prescribed through the target mass profile, rather than predicted by a
microscopic conversion law. We give both the tangential and averaged-pressure
versions and identify their different degeneracy conditions.

Our main results concern the conditions under which this construction has a
physical interpretation. We distinguish a zero of one component from an
algebraically degenerate split, state the additional information required for
a covariant exchange vector, and derive a positivity criterion. An explicit
finite-density example admits inner and outer trapping horizons and a
nonnegative decomposition into vacuum-like and traceless anisotropic sectors.
Its total source satisfies the null, weak, and dominant energy conditions.
We also examine a homogeneous cosmological counterpart, where the same
positivity obstruction prevents ordinary matter and radiation alone from
reproducing accelerated expansion.

For observational contact, we compute the photon sphere and shadow of the
stationary endpoint and compare the resulting fractional shadow size with
published Sagittarius A* constraints \cite{EHTSgrA2022VI,Vagnozzi:2022moj}.
This comparison probes the model's nonvacuum exterior. It does not provide
evidence for matter conversion behind a horizon or determine the radial
exchange function independently of the metric.

Sections~\ref{sec:husain} and \ref{sec:kiselev} develop the reconstruction.
Sections~\ref{sec:regularBH} and \ref{sec:horizon} treat the finite-density
example and its trapping horizons. Sections~\ref{sec:phases} and
\ref{sec:cosmology} discuss the phase profiles and cosmological counterpart.
Section~\ref{sec:numerics} presents the numerical checks, and
section~\ref{sec:observations} gives the observational comparison.
\end{revision}

\section{Generalized Vaidya sources and radial reconstruction}
\label{sec:husain}
\begin{revision}
The reconstruction starts from a prescribed metric and asks how its total
source can be partitioned into effective sectors. It does not supply an
initial-value problem for a microscopic collapsing medium. The null flux and
radial stresses must both be specified before the sector interaction can be
interpreted covariantly.
\end{revision}

\subsection{Dynamical geometry and effective fluid variables}
\label{subsec:dynamical_geometry}

 We consider a dynamical, spherically symmetric geometry in advanced
Eddington--Finkelstein coordinates,
\begin{equation}
ds^2
=
-\left(1-\frac{2M(v,r)}{r}\right)dv^2
+2\,dv\,dr
+r^2d\Omega^2 ,
\label{eq:gv_metric}
\end{equation}
where $d \Omega^2 = d \theta^2 + \sin^2 \theta \, d \phi^2$, \(v\) is an advanced null coordinate and \(M(v,r)\) is the
Misner--Sharp mass function. This generalized Vaidya, or Husain, geometry is
well suited to dynamical configurations sourced by anisotropic matter and a
null energy flux. We use units \(c=8\pi G=1\) , signature \((-+++)\), and
denote derivatives with respect to \(r\) and \(v\) by a prime and a dot,
respectively.

\begin{revision}
 The Einstein equations give the Hawking--Ellis Type-II effective variables
\begin{equation}
\rho=\frac{2M'}{r^2},
\qquad
P_t=-\frac{M''}{r},
\qquad
\sigma=\frac{2\dot M}{r^2},
\label{eq:effective_variables}
\end{equation}
where $\rho$ is the density eigenvalue of the non-null anisotropic sector, $P_t$ its tangential pressure, and
\(\sigma\) the null-flux density. The radial pressure is fixed by
 \begin{equation}
P_r=-\rho .
\label{eq:Pr_minus_rho}
\end{equation}
 The source is therefore anisotropic and should not be interpreted as an
isotropic perfect fluid.
\end{revision}

Covariant conservation of the total stress tensor, \(\nabla_\nu T^{\mu\nu}=0\), implies the radial balance equation
\begin{equation}
 r\rho'+2\rho+2P_t=0 .
\label{eq:radial_conservation}
\end{equation} 
 Substitution of Eq.~\eqref{eq:effective_variables} into
Eq.~\eqref{eq:radial_conservation} verifies this identity. It will be used
below to construct interacting effective sectors whose sum reproduces a
prescribed target geometry.

\subsection{Two effective anisotropic sectors}
\label{subsec:interacting_split}
\begin{revision}
We first choose two sectors with tangential pressures
\begin{equation}
P_b=\alpha\rho_b,\qquad P_{\rm rad}=\frac13\rho_{\rm rad}.
\end{equation}
The labels $b$ and ${\rm rad}$ retain the notation of the radial
reconstruction, but do not identify ordinary baryons or an isotropic photon
gas. Each non-null sector has $P_{r,i}=-\rho_i$. Even for
$0\leq\alpha\leq1$, the first is anisotropic. The second has trace
$-\rho_{\rm rad}+P_{r,\rm rad}+2P_{\rm rad}=-4\rho_{\rm rad}/3$;
it is not a traceless radiation source. Section~\ref{sec:kiselev} treats the
traceless anisotropic alternative.
\end{revision}

\begin{revision}
The second effective-sector density
\(\rho_{\rm rad}\) differs from the null flux \(\sigma\) defined in
Eq.~\eqref{eq:effective_variables}. The latter is fixed geometrically by the
advanced-time dependence of the mass function \(M(v,r)\), whereas
\(\rho_{\rm rad}\) denotes an effective matter component entering the radial
conservation equation. These quantities need not
be identified unless an additional closure relation is imposed.
\end{revision}

The total energy density and tangential pressure are taken to be
\begin{equation}
\rho_{\rm tot}=\rho_b+\rho_{\rm rad},
\qquad
P_{\rm tot}=P_b+P_{\rm rad}.
\end{equation}
\begin{revision}
We parameterize the departure from separate radial balance by
\end{revision}

\begin{revision}
\begin{align}
r\rho_{\rm rad}' +2\rho_{\rm rad}+2P_{\rm rad}
&=\beta(v,r)\rho_{\rm rad}, \nonumber\\
r\rho_b' +2\rho_b+2P_b
&=-\beta(v,r)\rho_{\rm rad}.
\label{eq:system}
\end{align}
Adding the two equations gives
\begin{equation}
r\rho_{\rm tot}'+2\rho_{\rm tot}+2P_{\rm tot}=0,
\end{equation}
which is precisely the radial conservation equation of the total Type-II
source. \rev{This verifies the radial projection of total conservation. Full
covariant conservation additionally includes the time-dependent flux
identity in section~\ref{subsec:covariant_exchange}.}
\end{revision}

Equivalently, one may write
\begin{equation}
T^{\mu\nu}=T^{\mu\nu}_{(b)}+T^{\mu\nu}_{({\rm rad})},
\qquad
\nabla_\nu T^{\mu\nu}=0,
\end{equation}
with
\begin{equation}
\nabla_\nu T^{\mu\nu}_{({\rm rad})}=Q^\mu,
\qquad
\nabla_\nu T^{\mu\nu}_{(b)}=-Q^\mu .
\end{equation}
\begin{revision}
Equation~\eqref{eq:system} fixes the radial balance of each sector; it does
not determine all components of $Q^\mu$. The total covariant conservation
law follows from the Einstein equations for the prescribed metric. To
promote the split to a specified covariant interaction one must additionally
choose the sector null fluxes, as shown below. The sign of $\beta$ is a sign
in a radial equation at fixed advanced time. It cannot establish the
direction of energy transfer in time. A divergence of $\beta$ also need not
imply a divergent unnormalized radial source $\beta\rho_{\rm rad}$.
\end{revision}

\begin{revision}
At fixed advanced time \(v\), Eq.~\eqref{eq:system} can be integrated
formally. The radiation equation gives
\begin{equation}
\rho_{\rm rad}(v,r)
=
\rho_{0r}(v)
\exp\left[
\int^r
\frac{\beta(v,\bar r)-\frac{8}{3}}{\bar r}
\,d\bar r
\right],
\label{eq:rho_rad_solution}
\end{equation}
where \(\rho_{0r}(v)\) is an integration function. The first-sector equation then
yields
\begin{strip}
\begin{equation}
\begin{aligned}
\rho_b(v,r)
={}&r^{-2(1+\alpha)}
\left[
C(v)
\right.\\[-1mm]
&\left.\quad-
\rho_{0r}(v)\int^r \bar r^{\,1+2\alpha}\beta(v,\bar r)
\exp\!\left(
\int^{\bar r}\frac{\beta(v,\tilde r)-\frac{8}{3}}{\tilde r}
\,d\tilde r\right)d\bar r
\right],
\end{aligned}
\label{eq:rho_b_solution}
\end{equation}
\end{strip}
where \(C(v)\) is fixed by boundary or initial data. In the absence of
interaction, \(\beta=0\), the second effective sector scales as
\(\rho_{\rm rad}\propto r^{-8/3}\). A nonzero \(\beta\) therefore modifies
this radial scaling and can either enhance or soften the growth of the
radiative component toward the center.
\end{revision}

\begin{revision}
We now introduce the target effective phase. This is not a third independently
conserved component. Instead, it denotes the total source
generated by the interacting two effective sectors:
\begin{equation}
\begin{aligned}
\rho_{\rm new}\equiv \rho_{\rm tot}
  &=\rho_b+\rho_{\rm rad},\\
P_{\rm new}\equiv P_{\rm tot}
  &=\alpha\rho_b+\frac{1}{3}\rho_{\rm rad}.
\end{aligned}
\label{eq:target_split}
\end{equation}
The inverse problem is then the following: given a desired effective phase
\((\rho_{\rm new},P_{\rm new})\), determine the radial reconstruction profile
\(\beta(v,r)\) required to realize it.
\end{revision}

\begin{revision}
Solving Eq.~\eqref{eq:target_split} algebraically gives
\begin{equation}
\begin{aligned}
\rho_{\rm rad}
  &=\frac{\alpha\rho_{\rm new}-P_{\rm new}}
          {\alpha-\frac{1}{3}},\\
\rho_b
  &=\frac{P_{\rm new}-\frac{1}{3}\rho_{\rm new}}
          {\alpha-\frac{1}{3}}.
\end{aligned}
\label{eq:algebraic_split}
\end{equation}
\rev{The split is unique for $\alpha\neq1/3$. At $\alpha=1/3$ a
solution exists only if $P_{\rm new}=\rho_{\rm new}/3$; when compatible,
it requires additional information to fix the partition.}
\end{revision}

\begin{revision}
Using the first equation in Eq.~\eqref{eq:system}, together with
\(P_{\rm rad}=\rho_{\rm rad}/3\), one obtains
\begin{equation}
\frac{\beta-\frac{8}{3}}{r}
=
\frac{\rho_{\rm rad}'}{\rho_{\rm rad}}
=
\frac{d}{dr}
\ln\left|\alpha\rho_{\rm new}-P_{\rm new}\right|.
\label{eq:log_beta_pressure}
\end{equation}
This already shows that \(\beta\) is fixed by the radial structure of the
target phase.
\end{revision}

\begin{revision}
An admissible target must satisfy the radial conservation law; its density
and tangential pressure cannot be chosen independently:
\end{revision}

\begin{equation}
r\rho_{\rm new}'+2\rho_{\rm new}+2P_{\rm new}=0,
\label{eq:new_conservation}
\end{equation}
then
\begin{equation}
P_{\rm new}
=
-\rho_{\rm new}
-\frac{r}{2}\rho_{\rm new}'.
\end{equation}
Consequently,
\begin{equation}
\begin{aligned}
\alpha\rho_{\rm new}-P_{\rm new}
&=(\alpha+1)\rho_{\rm new}
  +\frac{r}{2}\rho_{\rm new}'\\
&=\frac{1}{2}\left[
  (2\alpha+2)\rho_{\rm new}+r\rho_{\rm new}'\right].
\end{aligned}
\end{equation}
Equation~\eqref{eq:log_beta_pressure} may therefore be written equivalently as
\begin{equation}
\begin{aligned}
\frac{\beta-\frac{8}{3}}{r}
={}&\frac{1}{(2\alpha+2)\rho_{\rm new}+r\rho_{\rm new}'}\\
&\times\frac{d}{dr}\left[
(2\alpha+2)\rho_{\rm new}+r\rho_{\rm new}'\right].
\end{aligned}
\label{eq:log_beta_density}
\end{equation}

\begin{revision}
Finally, eliminating \(\rho_{\rm new}'\) by means of
Eq.~\eqref{eq:new_conservation}, Eq.~\eqref{eq:log_beta_pressure} gives \cite{Vertogradov:2025collapse}
\begin{equation}
\beta(v,r)
=
\frac{
\frac{2\alpha}{3}\rho_{\rm new}
-
\left(2\alpha+\frac{8}{3}\right)P_{\rm new}
-
rP_{\rm new}'
}{\alpha\rho_{\rm new}-P_{\rm new}}.
\label{eq:beta1}
\end{equation}
This is the central reconstruction formula of the present subsection. It
determines the effective radial reconstruction profile required for the interacting
two-sector system to mimic a prescribed, separately conserved target
phase \((\rho_{\rm new},P_{\rm new})\).
\end{revision}

The same expression can be written in terms of \(\rho_{\rm rad}\) using
Eq.~\eqref{eq:algebraic_split}:
\begin{equation}
\alpha\rho_{\rm new}-P_{\rm new}
=
\left(\alpha-\frac{1}{3}\right)\rho_{\rm rad},
\end{equation}
and therefore
\begin{equation}
\beta(v,r)
=
\frac{
\frac{2\alpha}{3}\rho_{\rm new}
-
\left(2\alpha+\frac{8}{3}\right)P_{\rm new}
-
rP_{\rm new}'
}{
\left(\alpha-\frac{1}{3}\right)\rho_{\rm rad}
}.
\end{equation}
\begin{revision}
For $\alpha\neq1/3$, a zero of $\alpha\rho_{\rm new}-P_{\rm new}$ is a
zero of the reconstructed second density. The algebraic decomposition
remains unique there. What fails is normalization of the radial source by
that density. This is different from $\alpha=1/3$, where the two equations
of state coincide and the algebraic map loses rank. Logarithmic derivatives
in this section are understood as derivatives of the absolute value on
intervals where the argument is nonzero.
\end{revision}

\subsection{Covariant exchange and the general two-sector map}
\label{subsec:covariant_exchange}
\begin{revision}
For two constant tangential coefficients $(\alpha,\chi)$, write
$P_b=\alpha\rho_b$, $P_X=\chi\rho_X$. The nondegenerate algebraic split is
\begin{equation}
\rho_X=\frac{\alpha\rho-P_t}{\alpha-\chi},\qquad
\rho_b=\frac{P_t-\chi\rho}{\alpha-\chi},\qquad \alpha\neq\chi.
\end{equation}
The second radial balance equation gives
\begin{equation}
\begin{aligned}
\beta_\chi
  &=2(1+\chi)+r\partial_r\ln|\rho_X|\\
  &=\frac{2\alpha\chi\rho-2(\alpha+\chi+1)P_t-rP_t'}
          {\alpha\rho-P_t}.
\end{aligned}
\label{eq:general_beta}
\end{equation}
It is defined where $\rho_X\neq0$. At $\alpha=\chi$ the algebraic equations
are compatible only if $P_t=\alpha\rho$, and the partition is then not unique.
At fixed $\alpha\neq\chi$, a zero of $\rho_X$ instead concerns the
normalization of its radial source.

A covariant completion can be stated explicitly. For each sector $i$,
\begin{equation}
T^\mu{}_{\nu(i)}=
\begin{pmatrix}
-\rho_i&0&0&0\\
\sigma_i&-\rho_i&0&0\\
0&0&P_i&0\\
0&0&0&P_i
\end{pmatrix},\qquad \sigma_b+\sigma_X=\sigma,
\end{equation}
in the coordinates $(v,r,\theta,\phi)$. Define
\begin{equation}
\mathcal A_i=r\rho_i'+2(\rho_i+P_i),\qquad
\mathcal J_i=-\dot\rho_i+\sigma_i'+\frac{2\sigma_i}{r}.
\end{equation}
Direct covariant differentiation yields
\begin{equation}
\begin{aligned}
Q_i^v&=-\frac{\mathcal A_i}{r},&
Q_i^r&=\mathcal J_i-F\frac{\mathcal A_i}{r},\\
Q_i^\theta&=Q_i^\phi=0.&&
\end{aligned}
\label{eq:full_exchange}
\end{equation}
Both $\mathcal A_{\rm tot}=0$ and $\mathcal J_{\rm tot}=0$ follow from
$\rho=2M'/r^2$, $P_t=-M''/r$, and $\sigma=2\dot M/r^2$.
The radial prescription $\mathcal A_X=\beta_\chi\rho_X$ determines only one
combination. The component flux $\sigma_X$ must also be chosen to specify
$Q_X^\mu$ on $r>0$. These equations expose the information absent from a
radial reconstruction alone.
\end{revision}

\subsection{Energy and convergence conditions for the Type-II source}
\label{subsec:energy_conditions_typeII}

The independent effective variables of the generalized Vaidya source are already given in Eqs.~\eqref{eq:effective_variables} and
\eqref{eq:Pr_minus_rho}. The radial combination
\(\rho+P_r\) is therefore identically zero, whereas the nontrivial angular
combination is \(\rho+P_t\). With the ingoing orientation adopted here,
non-negative null flux requires
\begin{equation}
 \sigma\geq0
\qquad\Longleftrightarrow\qquad
\dot M\geq0 .
\label{eq:positive_flux}
\end{equation} 

It is useful to distinguish carefully between the energy conditions imposed on
the effective stress tensor and the geometrical convergence conditions. In
Einstein gravity , the null energy condition (NEC) 
 is equivalent to the null
convergence condition (NCC). For the present Hawking--Ellis Type-II source,
the NEC/NCC requires
\begin{equation}
 \sigma\geq0,
\qquad
\rho+P_t\geq0,
\end{equation} 
or, in terms of the Misner--Sharp mass,
\begin{equation}
 \dot M\geq0,
\qquad
2M'-rM''\geq0 .
\label{eq:ncc_mass}
\end{equation} 
The weak energy condition (WEC) additionally requires non-negative energy
density,
\begin{equation}
 \rho\geq0
\qquad\Longleftrightarrow\qquad
M'\geq0 .
\end{equation}
 Consequently, the complete WEC conditions are
\begin{equation}
 \dot M\geq0,
\qquad
M'\geq0,
\qquad
2M'-rM''\geq0 .
\label{eq:wec_mass}
\end{equation}
 This explains why \(M'\geq0\) appears in our WEC test but not in a purely NCC
analysis (see Ref. \cite{Borissova:2025msp}): positivity of the local energy density is not part of the NCC .

The dominant energy condition (DEC) further requires the tangential stress not
to exceed the energy density in magnitude,
\begin{equation}
 \rho\geq |P_t|,
\end{equation} 
which is equivalent to
\begin{equation}
 \dot M\geq0,
\qquad
2M'\geq r|M''| .
\label{eq:dec_mass}
\end{equation} 
The inequality in Eq.~\eqref{eq:dec_mass} already implies \(M'\geq0\) and the NEC inequality .

The strong energy condition (SEC) is equivalent, through the Einstein
equations, to the timelike convergence condition (TCC). In addition to the
NCC inequalities, it requires
\begin{equation}
 \rho+P_r+2P_t=2P_t\geq0
\qquad\Longleftrightarrow\qquad
M''\leq0 .
\end{equation} 

Thus the TCC conditions are
\begin{equation}
 \dot M\geq0,
\qquad
2M'-rM''\geq0,
\qquad
-M''\geq0 .
\label{eq:tcc_mass}
\end{equation} 

 \begin{revision}
The WEC and TCC impose different additional conditions: $M'\geq0$ for the
former and $M''\leq0$ for the latter. Both also require the NCC inequalities
\cite{Borissova:2025hmj}.
\end{revision}

\begin{revision}
For a finite-curvature de~Sitter-like center,
\begin{equation}
 M(v,r)=M_3(v)r^3+O(r^4),
\qquad
M_3(v)>0,
\label{eq:regular_center_expansion}
\end{equation} 
one has
\begin{equation}
\begin{aligned}
\rho(v,0)&=6M_3(v),\\
P_t(v,0)&=-6M_3(v)=-\rho(v,0).
\end{aligned}
\label{eq:regular_center_state}
\end{equation} 
The WEC can therefore hold and the NCC is saturated at leading order, but the
TCC is violated in a neighborhood of the center because
\(M''\simeq6M_3r>0\). This distinction is essential for the singularity
theorems. Stationary de~Sitter-core regular black holes may satisfy the NCC
while evading the Penrose theorem through an inner Cauchy horizon and the
associated failure of global hyperbolicity. They evade the Hawking--Penrose
theorem through the local TCC violation expressed by \(M''>0\)
\cite{Borissova:2025msp,Borissova:2025hmj} .
\end{revision}

In a dynamical spacetime, satisfying Eqs.~\eqref{eq:ncc_mass} and
\eqref{eq:wec_mass} is only a local statement. It does not prove that the
maximal extension is globally hyperbolic or geodesically complete. Accordingly,
the explicit construction below should be interpreted as a locally regular
collapse model with apparent-horizon formation. Establishing a globally
nonsingular endpoint requires a separate analysis of null geodesics, the inner
horizon, and the global causal structure.

 \section{Kiselev-like model with averaged pressure}
\label{sec:kiselev}

\begin{revision}
The construction above used a tangential barotropic relation
\(P_b=\alpha\rho_b\) for the first effective component of the Type-II source.
For anisotropic fluids, however, it is often useful to characterize the matter
content through the averaged pressure
\[
\bar P=\frac{1}{3}(P_r+2P_t),
\]
rather than through the tangential pressure alone. This is the convention used
in Kiselev-type descriptions of anisotropic matter surrounding compact objects
\cite{Kiselev:2002bh,Visser2020Kiselev}. In the present section we reformulate the interacting
split in this averaged-pressure parametrization and derive the corresponding
radial reconstruction function.
\end{revision}

\subsection{Averaged barotropic equation of state}

\begin{revision}
For the generalized Vaidya source considered here, Eq.~\eqref{eq:Pr_minus_rho} fixes the radial pressure. Therefore the averaged
pressure is
\begin{equation}
\bar P
=
\frac{1}{3}\left(-\rho+2P_t\right).
\end{equation}
If the averaged pressure of the first anisotropic sector obeys
\begin{equation}
\bar P_b=\omega\rho_b,
\end{equation}
then the corresponding tangential pressure is
\begin{equation}
P_b
=
\frac{1}{2}(3\omega+1)\rho_b .
\end{equation}
It is convenient to write this again as
\begin{equation}
P_b=\alpha\rho_b,
\end{equation}
with
\begin{equation}
\alpha=\frac{3\omega+1}{2},
\qquad
\omega=\frac{2\alpha-1}{3}.
\label{eq:connection}
\end{equation}
Thus \(\alpha\) in this section denotes the tangential-pressure coefficient
associated with the averaged-pressure parameter \(\omega\). It should not be
confused with the equation-of-state parameter of an isotropic perfect fluid.
\end{revision}

\begin{revision}
A radiation-like averaged equation of state corresponds to
\begin{equation}
\omega_{\rm rad}=\frac{1}{3}.
\end{equation}
Because \(P_r=-\rho\) for the Type-II source, this gives
\begin{equation}
P_{\rm rad}=\rho_{\rm rad},
\end{equation}
rather than the isotropic FRW relation \(P=\rho/3\). This difference is the
origin of the different numerical coefficients appearing below. If one imposes
the tangential DEC on the first anisotropic sector, one
should restrict \(\alpha\leq1\), equivalently \(\omega\leq1/3\). More general
Kiselev-like effective sources may formally allow larger values of \(\alpha\),
but then the DEC must be checked separately.
\end{revision}

\subsection{Transition equations and reconstruction formula}

\begin{revision}
The interacting radial balance equations take the same form as before,
\begin{align}
r\rho_{\rm rad}' +2\rho_{\rm rad}+2P_{\rm rad}
&=\beta(v,r)\rho_{\rm rad},
\nonumber\\
r\rho_b' +2\rho_b+2P_b
&=-\beta(v,r)\rho_{\rm rad}.
\label{eq:kiselev_system}
\end{align}
In the averaged-pressure parametrization, the radiation-like component obeys
\(P_{\rm rad}=\rho_{\rm rad}\). Hence the first equation becomes
\begin{equation}
r\rho_{\rm rad}'+4\rho_{\rm rad}
=
\beta(v,r)\rho_{\rm rad}.
\end{equation}
At fixed advanced time \(v\), this gives
\begin{equation}
\rho_{\rm rad}(v,r)
=
\rho_{0r}(v)
\exp\left[
\int^r
\frac{\beta(v,\bar r)-4}{\bar r}\,d\bar r
\right].
\label{eq:kiselev_rhorad}
\end{equation}
The first-sector density is then
\begin{strip}
\begin{equation}
\begin{aligned}
\rho_b(v,r)
={}&r^{-2(1+\alpha)}
\left[
C(v)
\right.\\[-1mm]
&\left.\quad-
\rho_{0r}(v)\int^r \bar r^{1+2\alpha}\beta(v,\bar r)
\exp\!\left(
\int^{\bar r}\frac{\beta(v,\tilde r)-4}{\tilde r}
\,d\tilde r\right)d\bar r
\right],
\end{aligned}
\label{eq:kiselev_rhob}
\end{equation}
\end{strip}
where \(C(v)\) and \(\rho_{0r}(v)\) are fixed by initial or boundary data.
\end{revision}

\begin{revision}
As in the previous section, the target effective phase is defined as the
total source,
\begin{equation}
\rho_{\rm new}
=
\rho_b+\rho_{\rm rad},
\qquad
P_{\rm new}
=
\alpha\rho_b+\rho_{\rm rad}.
\label{eq:kiselev_target}
\end{equation}
This is not a third independently conserved component. It is the effective
source that the interacting two-sector split is required to mimic.
\end{revision}

\begin{revision}
Solving Eq.~\eqref{eq:kiselev_target} algebraically gives
\begin{equation}
\rho_{\rm rad}
=
\frac{\alpha\rho_{\rm new}-P_{\rm new}}{\alpha-1},
\qquad
\rho_b
=
\frac{P_{\rm new}-\rho_{\rm new}}{\alpha-1}.
\label{eq:kiselev_split}
\end{equation}
\rev{The averaged-pressure split is unique for $\alpha\neq1$.
At $\alpha=1$ it exists only if $P_{\rm new}=\rho_{\rm new}$, and the
partition is then undetermined.}
\end{revision}

\begin{revision}
Using Eq.~\eqref{eq:kiselev_rhorad}, one obtains
\begin{equation}
\frac{\beta-4}{r}
=
\frac{\rho_{\rm rad}'}{\rho_{\rm rad}}
=
\frac{d}{dr}
\ln\left|\alpha\rho_{\rm new}-P_{\rm new}\right|.
\label{eq:kiselev_log}
\end{equation}
If the target phase is separately required to satisfy the Type-II radial
conservation law 
 \eqref{eq:new_conservation}, then
Eq.~\eqref{eq:kiselev_log} yields
\begin{equation}
\beta(v,r)
=
\frac{
2\alpha\rho_{\rm new}
-
(2\alpha+4)P_{\rm new}
-
rP_{\rm new}'
}{\alpha\rho_{\rm new}-P_{\rm new}}.
\label{eq:beta}
\end{equation}
This is the averaged-pressure counterpart of Eq.~\eqref{eq:beta1}. It has the
same structural dependence on
\((\rho_{\rm new},P_{\rm new},P_{\rm new}')\), but the numerical coefficients
differ because the radiation-like component is defined by
\(\bar P_{\rm rad}=\rho_{\rm rad}/3\), which corresponds to
\(P_{\rm rad}=\rho_{\rm rad}\) in the tangential Type-II sector.
\end{revision}

Using Eq.~\eqref{eq:kiselev_split}, the denominator may be written as
\begin{equation}
\alpha\rho_{\rm new}-P_{\rm new}
=
(\alpha-1)\rho_{\rm rad},
\end{equation}
so that
\begin{equation}
\beta(v,r)
=
\frac{
2\alpha\rho_{\rm new}
-
(2\alpha+4)P_{\rm new}
-
rP_{\rm new}'
}{(\alpha-1)\rho_{\rm rad}}.
\end{equation}
\begin{revision}
The actual algebraic degeneracy is $\alpha=1$. At fixed $\alpha\neq1$, a
zero of $\alpha\rho_{\rm new}-P_{\rm new}$ instead makes the normalized
radial function undefined while the sector densities remain well defined.
The unnormalized radial balance and the curvature must be checked separately.
\end{revision}

\begin{revision}
In the explicit phase examples below, when Eq.~\eqref{eq:beta} is used, the
resulting \(\beta(v,r)\) should therefore be understood as an
averaged-pressure, Kiselev-like radial reconstruction profile. The original Husain
formula, Eq.~\eqref{eq:beta1}, gives the same type of reconstruction but with
the coefficients appropriate to the tangential radiation relation
\(P_{\rm rad}=\rho_{\rm rad}/3\).
\end{revision}

\section{Regular black holes from targeted interior profiles}
\label{sec:regularBH}

In this section we apply the reconstruction formalism to regular target geometries. We first show how a prescribed target density determines the Misner--Sharp mass and tangential pressure and state the corresponding local regularity conditions. We then construct an explicit finite-density dynamical example, determine the critical condition for trapping-horizon formation, and examine its energy and convergence conditions. The analysis establishes local curvature regularity and apparent-horizon formation, while questions concerning the maximal extension, geodesic completeness, and inner-horizon stability remain outside the scope of the present construction.

\begin{revision}
As established in Eqs.~\eqref{eq:regular_center_expansion} and
\eqref{eq:regular_center_state}, a finite-curvature center requires \(M(v,r)=O(r^3)\) and approaches the de~Sitter-like relation
\(P_t=-\rho\). The regularization problem can therefore be formulated as an
inverse reconstruction: one specifies a target effective phase
\((\rho_{\rm new},P_{\rm new})\), or equivalently a regular mass function
\(M(v,r)\), and then determines the radial reconstruction profile \(\beta(v,r)\) needed
for the interacting two-sector split to realize it.
\end{revision}

\subsection{Mass-profile reconstruction}

\begin{revision}
The target density determines the mass function through
\begin{equation}
M(v,r)=M(v,0)
+\frac{1}{2}\int_0^r
\rho_{\rm new}(v,\bar r)\,\bar r^2\,d\bar r .
\label{eq:mass_reconstruction}
\end{equation}
Regularity at the origin requires \(M(v,0)=0\). Once \(M(v,r)\) is known, the
corresponding tangential pressure is fixed by the field equation
\begin{equation}
P_{\rm new}(v,r)=-\frac{M''(v,r)}{r}.
\label{eq:pressure_from_mass}
\end{equation}
Equivalently, a separately conserved Type-II 
 target obeys
Eq.~\eqref{eq:new_conservation}. Substitution of
\((\rho_{\rm new},P_{\rm new})\) into the reconstruction
formula \(\eqref{eq:beta1}\), or into its averaged-pressure counterpart
\(\eqref{eq:beta}\), gives the radial reconstruction function required to support that
target geometry.
\end{revision}

\begin{revision}
Standard regular-black-hole interiors correspond to smooth mass profiles
which behave as
\begin{equation}
M(v,r)\sim r^3
\qquad
(r\to0),
\end{equation}
and approach a finite total mass at large radius. Dymnikova-type profiles
describe a rapid transition from an approximately constant-density core to a
dilute exterior \cite{Dymnikova:1992ux}. Bardeen-type profiles provide a
magnetically motivated regularization with the same \(r^3\) behavior at the
origin \cite{Bardeen:1968}. Hayward-type profiles implement a minimal-length
regularization and also interpolate between a de Sitter-like core and a
Schwarzschild-like exterior \cite{Hayward:2005gi}. In the present paper,
these profiles are interpreted as possible target geometries for which the
function $\beta(v,r)$ determines the radial source partition for each prescribed target geometry.
\end{revision}

Regularity does not require the SEC to hold everywhere.
On the contrary, a de Sitter-like core has \(P_t\simeq-\rho\), and therefore
violates the SEC in the central region. The relevant
physical requirement is that the violation be localized to the regularizing
core, while the NEC and WEC remain satisfied in the
physically relevant domain. The examples below show explicitly how this can
be achieved.

\subsection{Explicit dynamical reconstruction with horizon formation}
\label{sec:explicit_dynamic}

\begin{revision}
We now give a fully explicit dynamical example. The purpose is to show, in a
single construction, how a prescribed regular target phase determines the mass
function, the radial reconstruction profile, the apparent-horizon evolution, and the
energy conditions.
\end{revision}

We choose the finite-density target profile
\begin{equation}
\rho_{\rm new}(v,r)
=
\frac{\rho_c(v)}{1+x^4},
\qquad
x\equiv\frac{r}{r_0},
\label{eq:explicit_density}
\end{equation}
where \(r_0\) sets the size of the regular core and \(\rho_c(v)\) is a
monotonically increasing central density during collapse. The corresponding
mass function is
\begin{equation}
M(v,r)
=
\frac{1}{2}
\int_0^r
\rho_{\rm new}(v,\bar r)\,\bar r^2\,d\bar r
=
\frac{\rho_c(v)r_0^3}{2}\,{\cal I}(x),
\label{eq:explicit_mass}
\end{equation}
where
\begin{equation}
{\cal I}(x)
=
\int_0^x
\frac{y^2}{1+y^4}\,dy .
\label{eq:I_def}
\end{equation}
An explicit primitive is
\begin{align}
{\cal I}(x)
=&
\frac{\sqrt{2}}{8}
\ln\left(
\frac{x^2-\sqrt{2}x+1}{x^2+\sqrt{2}x+1}
\right)
\nonumber\\
&+\frac{\sqrt{2}}{4}\arctan(\sqrt{2}x-1)
\nonumber\\
&+\frac{\sqrt{2}}{4}\arctan(\sqrt{2}x+1).
\label{eq:I_explicit}
\end{align}
The total asymptotic mass is finite:
\begin{equation}
M_\infty(v)
=
\lim_{r\to\infty}M(v,r)
=
\frac{\pi}{4\sqrt{2}}\rho_c(v)r_0^3 .
\label{eq:explicit_total_mass}
\end{equation}

\begin{revision}
The tangential pressure follows either from
\(P_{\rm new}=-M''/r\) or from the Type-II conservation equation. One finds
\begin{equation}
P_{\rm new}(v,r)
=
\rho_c(v)
\frac{x^4-1}{(1+x^4)^2}.
\label{eq:explicit_pressure}
\end{equation}
Near the center,
\begin{equation}
\rho_{\rm new}(v,0)=\rho_c(v),
\qquad
P_{\rm new}(v,0)=-\rho_c(v),
\end{equation}
and
\begin{equation}
M(v,r)
=
\frac{\rho_c(v)}{6}r^3
+O(r^7).
\label{eq:explicit_core_expansion}
\end{equation}
Thus \(M/r^3\), \(M'/r^2\), and \(M''/r\) remain finite as \(r\to0\), and the
central curvature has the de Sitter limiting values.
\end{revision}

The null flux associated with the dynamical mass function is
\begin{equation}
\sigma(v,r)
=
\frac{2\dot M}{r^2}
=
\dot\rho_c(v)\,r_0\,
\frac{{\cal I}(x)}{x^2}.
\label{eq:explicit_sigma}
\end{equation}
Therefore \(\sigma\ge0\) whenever \(\dot\rho_c(v)\ge0\). A monotonically
increasing \(\rho_c(v)\) corresponds to positive ingoing null energy flux.

Substituting Eqs.~\eqref{eq:explicit_density} and
\eqref{eq:explicit_pressure} into the Husain reconstruction formula gives
\begin{equation}
\beta_{\rm H}(v,r)
=
\frac{
4\left[
(1-\alpha)x^8
+
(\alpha-9)x^4
+
2(\alpha+1)
\right]
}{
3(1+x^4)
\left[
(\alpha+1)+(\alpha-1)x^4
\right]
}.
\label{eq:explicit_beta}
\end{equation}
For example, for \(\alpha=1\),
\begin{equation}
\beta_{\rm H}(x)
=
\frac{8(1-2x^4)}{3(1+x^4)}.
\label{eq:explicit_beta_alpha1}
\end{equation}
\begin{revision}
The sign change of $\beta_{\rm H}$ describes the radial balance of this
chosen split. It has no independent implication for curvature regularity or
the direction of energy transfer in time.
\end{revision}

The apparent horizons are determined by
\begin{equation}
F(v,r_h)
=
1-\frac{2M(v,r_h)}{r_h}
=
0.
\label{eq:horizon_condition_F}
\end{equation}
Using Eq.~\eqref{eq:explicit_mass}, this becomes
\begin{equation}
\rho_c(v)r_0^2
\frac{{\cal I}(x_h)}{x_h}
=
1,
\qquad
x_h\equiv\frac{r_h}{r_0}.
\label{eq:horizon_condition_x}
\end{equation}
The function \({\cal I}(x)/x\) has a maximum at
\begin{equation}
x_{\rm crit}\simeq1.679,
\qquad
\left[\frac{{\cal I}(x)}{x}\right]_{\rm max}
\simeq0.31516.
\label{eq:xcrit}
\end{equation}
Two distinct marginal spheres, enclosing a trapped interval, exist when
\begin{equation}
\rho_c(v)r_0^2
>
\frac{1}{0.31516}
\simeq3.173 .
\label{eq:horizon_threshold}
\end{equation}
Thus the configuration is initially regular and horizonless, and develops a
locally regular trapped region only after the central density exceeds the
threshold \(\rho_c r_0^2\simeq3.173\).

\begin{revision}
A smooth accretion history in advanced time is
\begin{equation}
\begin{aligned}
\rho_c(v)&=\frac{\rho_\infty}{1+e^{-(v-v_0)/\tau}},\\
\tau&>0,\qquad \rho_\infty r_0^2>\lambda_c.
\end{aligned}
\label{eq:rho_c_profile}
\end{equation}
It approaches a dilute horizonless configuration as $v\to-\infty$ and has
$\dot\rho_c\geq0$. The first marginal sphere occurs at
\begin{equation}
\begin{aligned}
v_{\rm form}
  &=v_0+\tau\ln\!\left(
    \frac{\lambda_c}{\rho_\infty r_0^2-\lambda_c}\right),\\
\lambda_c&=3.172997024\ldots .
\end{aligned}
\label{eq:vform}
\end{equation}
For $\rho_\infty r_0^2=5$ this is $v_0+0.552\,\tau$.
This smooth time profile avoids the jump in $\dot\rho_c$ at the onset of a
Heaviside-switched exponential. Trapped spheres exist for $v>v_{\rm form}$;
at equality there is a single degenerate marginal sphere.
\end{revision}

At late times, the two apparent horizons are located at
\begin{equation}
r_-\simeq0.850\,r_0,
\qquad
r_+\simeq4.424\,r_0 .
\label{eq:late_horizons}
\end{equation}
This example therefore describes the local dynamical transition from a
regular horizonless configuration to a trapped geometry with a locally regular
de~Sitter-like core. The existence of two apparent horizons does not by itself
establish that the maximal spacetime extension is geodesically complete.

\begin{revision}
\begin{figure}[t]
\centering
\includegraphics[width=\linewidth]{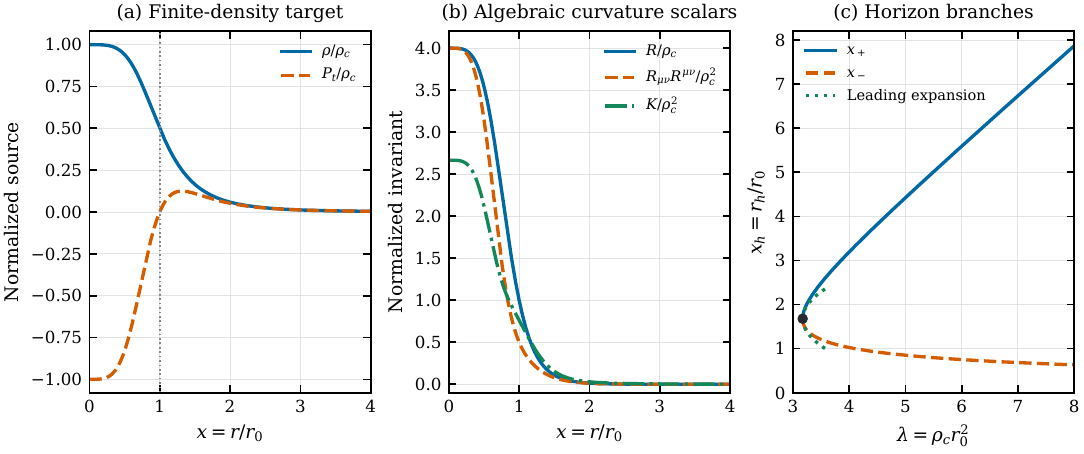}
\caption{
Radial structure, curvature regularity, and apparent-horizon formation for
the finite-density target
\(\rho=\rho_c/(1+x^4)\), with \(x=r/r_0\).
Panel (a) shows the normalized density and tangential pressure. The dotted
vertical line at \(x=1\) marks the change of sign of \(P_t\) and the outer
boundary of the de~Sitter-like TCC/SEC-violating core.
Panel (b) displays the exact Ricci scalar, Ricci-tensor square, and
Kretschmann scalar, normalized by the appropriate powers of \(\rho_c\).
All invariants remain finite at the center and tend to zero at spatial infinity.
Panel (c) shows the exact inner and outer apparent-horizon branches as
functions of \(\lambda=\rho_c r_0^2\). The branches meet at the critical
point \((\lambda_c,x_c)=(3.172997\ldots,1.678736\ldots)\); the dotted curves
represent the near-critical square-root expansion and demonstrate the
saddle-node character of horizon formation.
}
\label{fig:explicit_profile_invariants}
\end{figure}
\end{revision}

Figure~\ref{fig:explicit_profile_invariants} summarizes the local regularity
and horizon structure of the finite-density target introduced in
Eq.~\eqref{eq:explicit_density}. Panel~(a) shows the normalized density and
tangential pressure. The density is finite at the center and decreases
monotonically toward zero, whereas the pressure approaches the de~Sitter
relation \(P_t=-\rho_c\) as \(x\to0\). The pressure changes sign at \(x=1\),
which also marks the boundary of the region in which the nontrivial
TCC/SEC condition is violated. Outside this core, \(P_t\) becomes positive,
reaches its maximum value \(P_t/\rho_c=1/8\) at
\(x=3^{1/4}\), and subsequently decreases to zero. The sign change of
\(P_t\) is therefore a smooth transition between the vacuum-like core and
the dilute exterior, rather than a geometric singularity.

\begin{revision}
The exact curvature invariants are displayed in
Fig.~\ref{fig:explicit_profile_invariants}(b). For the target under
consideration, they can be written as
\begin{align}
\frac{R}{\rho_c}
&=
\frac{4}{(1+x^4)^2},
\label{eq:explicit_R}
\\
\frac{R_{\mu\nu}R^{\mu\nu}}{\rho_c^2}
&=
\frac{4(1+x^8)}{(1+x^4)^4},
\label{eq:explicit_Ricci2}
\\
\frac{K}{\rho_c^2}
&=
\frac{12{\cal I}^2(x)}{x^6}
\nonumber\\
&\quad-\frac{8{\cal I}(x)(1+3x^4)}
{x^3(1+x^4)^2}
\nonumber\\
&\quad+\frac{4(1+2x^4+5x^8)}
{(1+x^4)^4},
\label{eq:explicit_K}
\end{align}
where \(K\equiv
R_{\mu\nu\rho\sigma}R^{\mu\nu\rho\sigma}\). Their central limits are
\begin{equation}
\begin{aligned}
R(0)&=4\rho_c,\\
R_{\mu\nu}R^{\mu\nu}(0)&=4\rho_c^2,\\
K(0)&=\frac{8}{3}\rho_c^2.
\end{aligned}
\label{eq:explicit_invariant_limits}
\end{equation}
All three invariants are finite throughout the displayed domain and decay
continuously in the exterior. In particular, the Ricci invariants vanish
more rapidly than the Kretschmann scalar, whose leading large-radius
behavior reflects the finite asymptotic Schwarzschild mass. This verifies finiteness of the displayed algebraic curvature invariants;
the differentiability of the central extension is addressed separately below.
\end{revision}

Panel~(c) presents the inner and outer roots of the apparent-horizon
equation as functions of
\(\lambda=\rho_c r_0^2\). No positive horizon exists for
\(\lambda<\lambda_c\), whereas at
\begin{equation}
\lambda_c=3.172997024\ldots,
\qquad
x_c=1.678736111\ldots,
\label{eq:explicit_critical_values}
\end{equation}
the two roots merge into a degenerate trapping horizon. For
\(\lambda>\lambda_c\), this root separates into an inner branch \(x_-\)
and an outer branch \(x_+\). Close to the threshold, the branches satisfy
\begin{equation}
x_{\pm}
=
x_c
\pm
1.069603087\,
\sqrt{\lambda-\lambda_c}
+
O(\lambda-\lambda_c),
\label{eq:explicit_saddle_node}
\end{equation}
in agreement with the dotted curves in
Fig.~\ref{fig:explicit_profile_invariants}(c). The square-root opening is
the characteristic behavior of a saddle-node bifurcation and shows that
the trapped region is created through the simultaneous formation of inner
and outer marginal surfaces.

Far above the threshold, the two branches probe different physical
scales. The small-radius expansion of the mass function gives
\(x_-\sim\sqrt{3/\lambda}\), so the inner horizon moves toward the regular
core as the central density increases. By contrast, the outer branch
approaches
\begin{equation}
x_+
\sim
\frac{\pi}{2\sqrt{2}}\,\lambda,
\end{equation}
which is the Schwarzschild-radius behavior associated with the finite
asymptotic mass. The two branches therefore distinguish the inner trapping horizon from the total-mass-controlled outer horizon.

\begin{revision}
\paragraph{Differentiability at the dynamical center.}
Bounded curvature invariants must be distinguished from smoothness. A regular
local time $T$ can be introduced near the center by solving
$\partial_r v|_T=1/F$ with $v(T,0)=T$. With $a(T)=\rho_c(T)/3$, one obtains
\begin{equation}
\begin{aligned}
v&=T+r+\frac{a(T)}3r^3+O(r^4),\\
F&=1-a(T)r^2-\dot a(T)r^3+O(r^4).
\end{aligned}
\end{equation}
The mixed radial-time metric term then vanishes. In local Cartesian spatial
coordinates the resulting metric admits a $C^2$ central extension for smooth
$\rho_c(v)$. During accretion it is generically not $C^3$: for example,
\begin{equation}
R(T,r)=4\rho_c(T)+4\dot\rho_c(T)r+O(r^2)
\end{equation}
is continuous but has a radial cusp when $\dot\rho_c\neq0$. The claim here
is finite continuous curvature, not regularity of all curvature derivatives.
The stationary endpoint has no such time-induced cusp.
\end{revision}

\subsection{Positivity of the component decomposition}
\label{subsec:component_positivity}

Regularity and the energy conditions of the total source do not guarantee
that a particular algebraic decomposition of that source has nonnegative
component densities. To make this distinction explicit, consider the
general two-sector split
\begin{equation}
\rho=\rho_b+\rho_X,
\qquad
P_t=\alpha\rho_b+\chi\rho_X,
\label{eq:general_component_split}
\end{equation}
where \(\alpha\) and \(\chi\) are the tangential equation-of-state
parameters of the two effective sectors. For \(\alpha\neq\chi\), the
component densities are
\begin{equation}
\rho_X
=
\frac{\alpha\rho-P_t}{\alpha-\chi},
\qquad
\rho_b
=
\frac{P_t-\chi\rho}{\alpha-\chi}.
\label{eq:general_component_densities}
\end{equation}
Introducing \(w_t=P_t/\rho\), their fractional contributions become
\begin{equation}
\frac{\rho_X}{\rho}
=
\frac{\alpha-w_t}{\alpha-\chi},
\qquad
\frac{\rho_b}{\rho}
=
\frac{w_t-\chi}{\alpha-\chi}.
\label{eq:component_fractions}
\end{equation}
For a positive total density, both component densities are nonnegative if
and only if
\begin{equation}
\min(\alpha,\chi)
\leq
w_t
\leq
\max(\alpha,\chi).
\label{eq:component_positivity_condition}
\end{equation}
This condition is independent of the regularity and energy conditions of
the total stress tensor.

\begin{revision}
For the explicit target,
\begin{equation}
w_t(x)=\frac{x^4-1}{x^4+1},
\end{equation}
and hence \(w_t\to-1\) at the de~Sitter center. The original Husain
decomposition with \(\alpha=1\) and \(\chi=1/3\) yields
\begin{equation}
\frac{\rho_b^{\rm H}}{\rho_c}
=
\frac{x^4-2}{(1+x^4)^2},
\qquad
\frac{\rho_{\rm rad}^{\rm H}}{\rho_c}
=
\frac{3}{(1+x^4)^2}.
\label{eq:signed_husain_components}
\end{equation}
Although these two contributions reproduce the correct total density and
pressure, the first component is negative for
\begin{equation}
0\leq x<2^{1/4}.
\label{eq:negative_baryonic_domain}
\end{equation}
This region is indicated by the shading in
Fig.~\ref{fig:vacuum_completion}(a). The negative component density is not
a curvature singularity and does not imply that the total source violates
the WEC. It instead shows that a positive baryonic--radiative
interpretation is incompatible with the vacuum-like tangential pressure
required at the regular center.
\end{revision}

\begin{revision}
A nonnegative-density decomposition is obtained by choosing a vacuum-like second
sector with \(\chi=-1\). For \(\alpha=1\), one then finds
\begin{equation}
\begin{aligned}
\frac{\rho_b}{\rho_c}&=\frac{x^4}{(1+x^4)^2},\\
\frac{\rho_X}{\rho_c}&=\frac{1}{(1+x^4)^2},
\qquad \chi=-1.
\end{aligned}
\label{eq:positive_vacuum_components}
\end{equation}
\end{revision}

\begin{revision}
\begin{figure}[t]
\centering
\includegraphics[width=\linewidth]{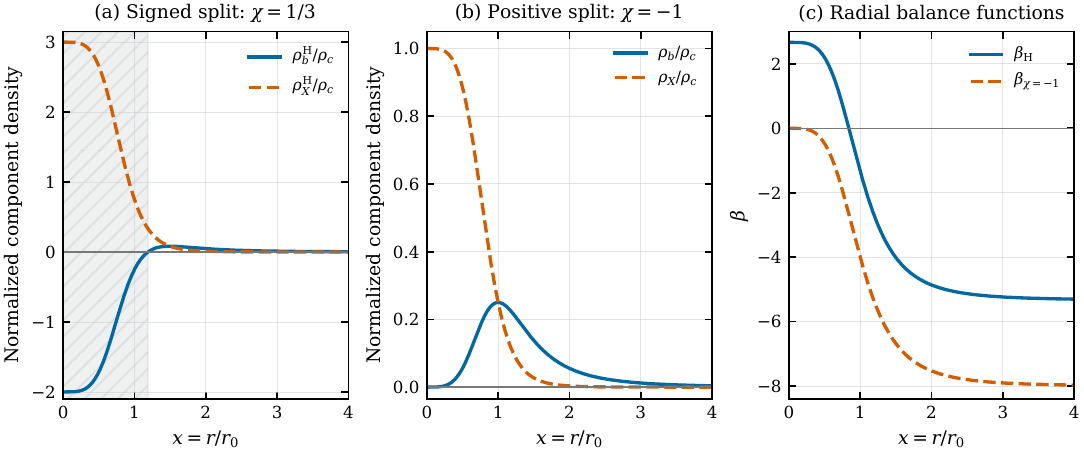}
\caption{
Comparison of two component decompositions of the explicit finite-density
target for \(\alpha=1\).
Panel (a) shows the original Husain split with
\(\chi=1/3\). Although the total source is regular and has positive density,
the reconstructed first component is negative in the shaded region
\(x<2^{1/4}\); the decomposition is therefore signed near the de~Sitter
core.
Panel (b) shows the vacuum-sector completion with \(\chi=-1\), for which
both component densities are nonnegative at every radius. The vacuum-like
sector dominates the center, whereas the first anisotropic sector becomes
dominant in the exterior.
Panel (c) compares the corresponding radial reconstruction functions.
Their signs characterize the radial balance of the chosen algebraic split
and should not be interpreted as microscopic conversion rates without an
additional covariant null-flux closure.
}
\label{fig:vacuum_completion}
\end{figure}
\end{revision}

\begin{revision}
Both densities are nonnegative for all \(x\geq0\), as shown in
Fig.~\ref{fig:vacuum_completion}(b). The vacuum-like sector supplies the
entire central density, while the first-sector density vanishes at the
origin, reaches its maximum near the transition region, and dominates over
the more rapidly decaying vacuum contribution in the exterior. Their
pressures combine to reproduce exactly the prescribed target pressure,
\begin{equation}
P_t=\rho_b-\rho_X
=
\rho_c\frac{x^4-1}{(1+x^4)^2}.
\end{equation}
The decomposition describes
of the radial transition from a vacuum-dominated regular core to a
matter-dominated exterior.
\end{revision}

The corresponding radial reconstruction profiles are
\begin{equation}
\beta_{\rm H}
=
\frac{8(1-2x^4)}{3(1+x^4)},
\qquad
\beta_{\chi=-1}
=
-\frac{8x^4}{1+x^4},
\label{eq:beta_husain_vacuum_comparison}
\end{equation}
and are compared in Fig.~\ref{fig:vacuum_completion}(c). The original
Husain profile begins at \(8/3\), changes sign at \(x=2^{-1/4}\), and
approaches \(-16/3\) asymptotically. The vacuum-sector profile vanishes at
the center and decreases monotonically toward \(-8\). These signs should
be interpreted only within the radial balance equations. In the absence
of an additional closure specifying the component null fluxes, neither
profile defines a unique covariant exchange four-vector or a microscopic
time-directed conversion rate. The principal result of the comparison is
therefore the existence of a globally nonnegative algebraic
decomposition, rather than the sign of the reconstruction function itself.
\begin{revision}
For the positive vacuum completion a specific covariant choice is
\begin{equation}
\begin{aligned}
\sigma_X&=0,& \sigma_b&=\sigma,\\
T_X^{\mu\nu}&=-\rho_Xg^{\mu\nu},&
Q_X^\mu&=-\nabla^\mu\rho_X.
\end{aligned}
\label{eq:vacuum_covariant_completion}
\end{equation}
Thus $Q_X^v=-\rho_X'$ and $Q_X^r=-\dot\rho_X-F\rho_X'$. This realizes the
radial function in equation~\eqref{eq:beta_husain_vacuum_comparison} and gives
an explicit exchange vector on $r>0$. The first sector has
$(P_{r,b},P_{t,b})=(-\rho_b,\rho_b)$ and is traceless; its label does not
make it baryonic matter. Although both densities are nonnegative, the
vacuum density inherits the central radial cusp during accretion, so this
particular exchange vector need not have a continuous direction-independent
limit at the center. A smooth microscopic two-sector completion remains an
additional requirement. This limitation does not alter conservation on $r>0$ or the
$C^2$ extension of the total metric.
\end{revision}

\subsection{Energy conditions for the explicit dynamical model}

For the target profile \eqref{eq:explicit_density}, the effective tangential
equation-of-state parameter is
\begin{equation}
w_{\rm eff}(x)
=
\frac{P_{\rm new}}{\rho_{\rm new}}
=
\frac{x^4-1}{x^4+1}.
\label{eq:weff_explicit}
\end{equation}
The density is non-negative,
\begin{equation}
\rho_{\rm new}
=
\frac{\rho_c}{1+x^4}
\ge0,
\end{equation}
and
\begin{equation}
\rho_{\rm new}+P_{\rm new}
=
\frac{2\rho_c x^4}{(1+x^4)^2}
\ge0 .
\label{eq:nec_explicit}
\end{equation}
 Together with the non-negative flux, the NCC/NEC and WEC are satisfied. The dominant energy
condition is also satisfied because
\begin{equation}
-1\leq w_{\rm eff}(x)\leq1,
\end{equation}
or equivalently
\begin{equation}
\rho_{\rm new}-|P_{\rm new}|\ge0 .
\label{eq:dec_explicit}
\end{equation}

The SEC contains the combination
\begin{equation}
\rho_{\rm new}+P_r+2P_t=2P_{\rm new},
\end{equation}
since \(P_r=-\rho_{\rm new}\). Therefore
\begin{equation}
\rho_{\rm new}+P_r+2P_t
=
\frac{2\rho_c(x^4-1)}{(1+x^4)^2}.
\label{eq:sec_explicit}
\end{equation}
The SEC is violated for \(x<1\), namely inside the compact
de Sitter-like core, and is restored for \(x>1\). This is precisely the desired local structure for a de~Sitter-core
regular geometry: the NCC/NEC and WEC are preserved, the DEC is preserved, and
the TCC/SEC violation is localized in the regularizing core. In particular,
\(P_{\rm new}<0\) for \(x<1\) implies \(M''>0\) there, in agreement with the
TCC analysis of Ref.~\cite{Borissova:2025hmj}.

\begin{revision}
The ratio of radial gradients is
\begin{equation}
\mathcal{S}_t(x)\equiv
\left.\frac{\partial_r P_t}{\partial_r\rho}\right|_v
=\frac{x^4-3}{x^4+1},
\label{eq:cs2_explicit}
\end{equation}
with central value defined by the limit. It is negative for $x<3^{1/4}$
and tends to unity at large radius. This derivative along a prescribed
background is not, by itself, a characteristic sound speed. A perturbation
closure for the radial stress, tangential stress, null flux, and sector
interaction is absent. Consequently, neither causal propagation nor a
short-wavelength instability follows from equation~\eqref{eq:cs2_explicit}.
\end{revision}

\section{Apparent horizons and causal evolution}
 \label{sec:horizon}

 Local curvature regularity does not by itself imply the existence of a black
hole; a trapped region must also form. In the generalized Vaidya geometry this
is monitored by
\begin{equation}
F(v,r)=1-\frac{2M(v,r)}{r}.
\label{eq:F_def}
\end{equation}
Marginally trapped sphere satisfies
\begin{equation}
F(v,r_h)=0,
\qquad
r_h(v)=2M\bigl(v,r_h(v)\bigr).
\label{eq:horizon_condition_general} \end{equation} 
The appearance of two roots is the local signature of the inner and outer
trapping horizons familiar from regular-black-hole geometries.

Differentiating Eq.~\eqref{eq:horizon_condition_general} with respect to the
advanced time gives
\begin{equation}
\dot r_h(v)
=
\frac{2\dot M(v,r_h)}{1-2M'(v,r_h)},
\label{eq:horizon_velocity}
\end{equation}
provided 
\(1-2M'(v,r_h)\neq0\). The numerator is proportional to the null flux defined in Eq.~\eqref{eq:effective_variables}. Hence positive ingoing flux
 tends to increase an outer trapping horizon when \(1-2M'(v,r_h)>0\). The limiting case
\(1-2M'(v,r_h)=0\) corresponds to a degenerate trapping horizon and must be
treated separately.

 On the horizon, \(F=0\), and a tangent to \(r=r_h(v)\) has norm
\begin{equation}
ds^2_{\rm hor}=2\dot r_h\,dv^2 .
\end{equation}

\begin{revision}
 The horizon world tube is therefore spacelike for \(\dot r_h>0\), null for
\(\dot r_h=0\), and timelike for 
 \(\dot r_h<0\). These statements provide
local checks of the trapping-horizon evolution .
\end{revision}

 The independent local consistency requirements have already been collected in
Sec.~\ref{subsec:energy_conditions_typeII}: regularity is governed by
Eq.~\eqref{eq:regular_center_expansion}, the NCC by
Eq.~\eqref{eq:ncc_mass}, the WEC by Eq.~\eqref{eq:wec_mass}, the DEC by
Eq.~\eqref{eq:dec_mass}, and the TCC by Eq.~\eqref{eq:tcc_mass}. Repeating
those inequalities here would obscure the distinct role of the horizon
analysis .

 Finally, apparent horizons are quasi-local objects. Their existence and local
regularity do not determine whether the inner horizon becomes a Cauchy horizon,
whether the maximal extension is globally hyperbolic, or whether all causal
geodesics are complete. If a stationary inner Cauchy horizon forms, the
Penrose theorem can be evaded through the failure of global hyperbolicity even
when the NCC is satisfied. \begin{revision}
If a noncompact Cauchy surface, a closed future-trapped surface, and the NCC
are retained, the Penrose theorem implies future null geodesic
incompleteness. The generic condition is not an assumption of that theorem
\cite{Penrose:1964wq}. Local TCC violation therefore does not resolve this
global obstruction. The present construction establishes finite local
curvature and trapping-horizon formation; it does not establish a complete
global extension or stability of an inner horizon.
\end{revision}

\section{Phase-structured target matter}
\label{sec:phases}

\begin{revision}
In the preceding sections the target source
\((\rho_{\rm new},P_{\rm new})\) was kept generic. We now illustrate how
different high-density equations of state are reflected in the reconstructed
radial reconstruction function \(\beta(v,r)\). The purpose of this section is not to
claim that every profile considered below gives a complete regular-black-hole
interior. Rather, the examples should be understood as illustrative target
phases: they show how polytropic matter, MIT-bag quark matter, and
condensate-inspired matter imprint different structures on the effective
radial reconstruction profile.
\end{revision}

\begin{revision}
Only target profiles satisfying
\begin{equation}
\rho_{\rm new}(v,r)\ge0,
\qquad
M(v,r)\sim r^3
\quad
(r\to0),
\end{equation}
can represent the positive-density regular centers considered here. Nonnegative density is a physical restriction rather than a mathematical necessity for finite curvature. Profiles that diverge at the origin or become
negative in part of their domain must instead be interpreted as local or
intermediate phases, to be matched to a regular inner core such as the one
constructed in Sec.~\ref{sec:explicit_dynamic}. In this section we use the
averaged-pressure reconstruction formula \(\eqref{eq:beta}\), unless stated
otherwise.
\end{revision}

\subsection{Polytropic target phase}

We first consider a target phase obeying the polytropic equation of state
\begin{equation}
P_p=\eta\rho_p^\gamma,
\label{eq:poly_eos}
\end{equation}
where \(\eta\) is the polytropic constant and
\begin{equation}
\gamma=1+\frac{1}{N}
\end{equation}
with \(N\) the polytropic index. Polytropic equations of state are widely used
as effective descriptions of compact-object matter when many-body interactions
and finite-density effects produce deviations from a strictly linear
barotropic relation.

For the Type-II source, the target phase must satisfy the radial conservation
law
\begin{equation}
r\rho_p'+2\rho_p+2P_p=0 .
\label{eq:poly_conservation}
\end{equation}
Using Eq.~\eqref{eq:poly_eos}, this gives
\begin{equation}
P_p'
=
-\frac{2\gamma}{r}\eta\rho_p^\gamma
-\frac{2\gamma}{r}\eta^2\rho_p^{2\gamma-1}.
\label{eq:poly_pprime}
\end{equation}
Equivalently, this expression follows by differentiating
\(P_p=\eta\rho_p^\gamma\) and using Eq.~\eqref{eq:poly_conservation}.

\begin{revision}
Identifying
\begin{equation}
\rho_{\rm new}=\rho_p,
\qquad
P_{\rm new}=P_p,
\end{equation}
the averaged-pressure reconstruction formula gives
\begin{equation}
\beta_p(v,r)
=
\frac{
2\alpha\rho_p
-
(2\alpha+4)P_p
-
rP_p'
}{\alpha\rho_p-P_p}.
\label{eq:beta_poly_general}
\end{equation}
Substituting Eqs.~\eqref{eq:poly_eos} and \eqref{eq:poly_pprime}, one obtains
\begin{equation}
\beta_p(v,r)
=
2+
\frac{
2\eta(\gamma-1-\alpha)\rho_p^\gamma
+
2\gamma\eta^2\rho_p^{2\gamma-1}
}{\alpha\rho_p-\eta\rho_p^\gamma}.
\label{eq:beta_poly}
\end{equation}
The first correction term measures the mismatch between the baryonic
tangential index \(\alpha\) and the effective polytropic response
\(\gamma-1\). The second term is nonlinear in the polytropic self-interaction
and becomes important at high density. A pole in Eq.~\eqref{eq:beta_poly}
occurs when
\begin{equation}
\alpha\rho_p=\eta\rho_p^\gamma,
\end{equation}
\rev{which is a zero of the second component at fixed $\alpha\neq1$.
The algebraic split remains unique; its normalized radial function has a
pole. This is distinct from a curvature singularity.} The geometric regularity must still be checked from
the reconstructed mass function.
\end{revision}

\begin{revision}
For comparison with exact Type-II polytropic solutions \cite{Vertogradov:2024grg}, one may write
\begin{equation}
M(v,r)=M_0(v)+\int W(v,r)\,dr,
\end{equation}
where
\begin{equation}
W(v,r)
=
\left[
D(v)-2^{\gamma-1}\eta r^{2-2\gamma}
\right]^{\frac{1}{1-\gamma}},
\end{equation}
with \(M_0(v)\) and \(D(v)\) fixed by boundary or matching conditions. Then
\(\rho_p=2W/r^2\), and Eq.~\eqref{eq:beta_poly} gives the corresponding
radial reconstruction profile. Whether this profile represents a regular core depends on
the small-\(r\) behavior of \(M(v,r)\); in particular, regularity requires
\(M(v,r)\sim r^3\).
\end{revision}

\begin{revision}
For $\eta>0$, $\gamma>1$, and $D>0$, the positive branch can be written as
\begin{equation}
\begin{aligned}
\rho_p&=\left[2^{1-\gamma}D\,r^{2(\gamma-1)}-\eta
       \right]^{-1/(\gamma-1)},\\
r&>r_{\min}=\left(\frac{2^{\gamma-1}\eta}{D}
       \right)^{1/[2(\gamma-1)]}.
\end{aligned}
\label{eq:poly_domain}
\end{equation}
Its density diverges as $r\downarrow r_{\min}$. It cannot approach a
positive finite-density center, since $P_p(0)=-\rho_p(0)$ conflicts with
$P_p=\eta\rho_p^\gamma>0$. The separate case $\gamma=1$ gives
$\rho_p=A(v)r^{-2(1+\eta)}$. A finite-layer application must be matched
before its singular boundary is reached; positivity on a plotted interval
alone does not establish a regular configuration. The polytropic relation
here applies to tangential stress rather than to an isotropic stellar fluid.
\end{revision}

\subsection{MIT-bag quark-matter layer}

\begin{revision}
We next impose a bag-model-inspired tangential pressure relation
\begin{equation}
P_q=\frac{1}{3}\left(\rho_q-4b\right),
\label{eq:mit_eos}
\end{equation}
where $b$ is a constant inspired by the MIT bag model
\cite{Witten:1984rs,Farhi:1984qu}. \rev{Here this relation is imposed on
$P_t$, while $P_r=-\rho$. It is therefore an anisotropic mathematical
analogue of the isotropic quark-matter equation of state, not a derivation
of deconfined matter dynamics.} Solving
the Type-II radial conservation law with this equation of state gives
\begin{equation}
\rho_q=C(v)r^{-8/3}+b,
\qquad
P_q=\frac{C(v)}{3}r^{-8/3}-b.
\label{eq:quark_rho_p}
\end{equation}
Here \(C(v)\) is an integration function. The pressure derivative is
\begin{equation}
rP_q'
=
-\frac{8C(v)}{9}r^{-8/3}.
\label{eq:quark_pprime}
\end{equation}
\end{revision}

Substitution into Eq.~\eqref{eq:beta} yields
\begin{equation}
\beta_q(v,r)
=
\frac{
2\alpha\rho_q
-
(2\alpha+4)P_q
-
rP_q'
}{\alpha\rho_q-P_q}.
\label{eq:beta_quark_general}
\end{equation}
After multiplying numerator and denominator by \(r^{8/3}\), this becomes
\begin{equation}
\beta_q(v,r)
=
\frac{
4\left[
(3\alpha-1)C(v)
+
9b(\alpha+1)r^{8/3}
\right]
}{
3\left[
(3\alpha-1)C(v)
+
3b(\alpha+1)r^{8/3}
\right]
}.
\label{eq:beta_quark}
\end{equation}

\begin{revision}
For $b>0$, $C(v)>0$, and $\alpha\neq1,1/3$, the limiting values are
\begin{equation}
\beta_q\longrightarrow\frac43\quad(r\to0),\qquad
\beta_q\longrightarrow4\quad(r\to\infty).
\end{equation}
For $0\leq\alpha<1/3$, a zero of the second density separates these limits:
\begin{equation}
r_\star^{8/3}=\frac{(1-3\alpha)C(v)}{3b(\alpha+1)}.
\end{equation}
The radial function has a pole there. For $\alpha=1/3$ and $b>0$, the
averaged-pressure split is nondegenerate and gives $\beta_q=4$; its
second density is $-2b$, so it is a signed decomposition. The value
$\alpha=1$ is excluded because the averaged-pressure sectors then have
identical tangential equations of state. A finite formal expression at
$\alpha=1$ does not cure that algebraic degeneracy. If $b=0$ and
$\alpha=1/3$, the second density vanishes identically and its normalized
radial function is undetermined.
\end{revision}

\begin{revision}
The profile \(\rho_q=C(v)r^{-8/3}+b\) is singular at the origin unless
\(C(v)=0\). Therefore it cannot by itself represent the central regular core
of a regular black hole. Its role here is more limited: it provides an
effective description of an intermediate effective layer or of a phase to be
matched to a regular inner profile. A complete regular-black-hole interior
must replace the central region by a density profile satisfying
\(M(v,r)\sim r^3\).
\end{revision}

\begin{revision}
The same bag-model profile tends to $\rho_q=b$ at large radius. Hence its
mass grows as $br^3/6$ and is not asymptotically finite. For $C=0$ it is a
de Sitter patch; for $C\neq0$ its central density also diverges. Using it
as a layer requires a specified matching construction at both ends.
\end{revision}

\subsection{Condensate-inspired target profile}

Finally, we consider a condensate-inspired density profile of the form
\begin{equation}
\rho_{\rm BEC}(r)
=
\frac{a}{kr}\sin(kr),
\label{eq:bec_density}
\end{equation}
where \(a\) and \(k\) are constants. This type of oscillatory profile is
motivated by Bose--Einstein-condensate models of compact objects and dark
matter halos \cite{Chavanis:2011uv,Harko:2011zt}. It is regular at the origin,
since
\begin{equation}
\frac{\sin(kr)}{kr}\to1
\qquad
(r\to0),
\end{equation}
but it changes sign after the first zero of \(\sin(kr)\). Hence it is
physically meaningful only in a finite domain where \(\rho_{\rm BEC}\ge0\), or
as part of a matched construction.

Using the Type-II conservation equation
\begin{equation}
r\rho_{\rm BEC}'+2\rho_{\rm BEC}+2P_{\rm BEC}=0,
\end{equation}
one obtains
\begin{equation}
P_{\rm BEC}
=
-\frac{a}{2}
\left[
\frac{\sin x}{x}
+
\cos x
\right],
\qquad
x\equiv kr.
\label{eq:bec_pressure}
\end{equation}
Moreover,
\begin{equation}
rP_{\rm BEC}'
=
\frac{a}{2}
\left[
\frac{\sin x}{x}
-
\cos x
+
x\sin x
\right].
\label{eq:bec_pprime}
\end{equation}
The density is
\begin{equation}
\rho_{\rm BEC}
=
a\frac{\sin x}{x}.
\end{equation}

\begin{revision}
Substitution into Eq.~\eqref{eq:beta} gives
\begin{equation}
\beta_{\rm BEC}(v,r)
=
\frac{
2\alpha\rho_{\rm BEC}
-
(2\alpha+4)P_{\rm BEC}
-
rP_{\rm BEC}'
}{\alpha\rho_{\rm BEC}-P_{\rm BEC}}.
\end{equation}
After simplification, one finds
\begin{equation}
\beta_{\rm BEC}(x)
=
\frac{\substack{(6\alpha+3)\sin x+(2\alpha+5)x\cos x\\
-x^2\sin x}}
{(2\alpha+1)\sin x+x\cos x}.
\label{eq:beta_bec}
\end{equation}
The denominator vanishes when
\begin{equation}
(2\alpha+1)\sin x+x\cos x=0.
\label{eq:bec_degeneracy}
\end{equation}
\rev{At such radii the second-component density vanishes. For
$\alpha\neq1$ its algebraic value remains unique, but the normalized
radial function is undefined. The total mass and curvature are regular
there; the pole is a property of the chosen normalization.}
\end{revision}

The BEC-inspired profile is regular at \(r=0\), but it cannot be used as a
global density profile without restriction, because it becomes negative after
the first node. Therefore we use it only as a local condensate-inspired
example, valid for \(0<kr<\pi\) or for a finite region matched to another
phase before \(\rho_{\rm BEC}\) changes sign. A complete regular-black-hole
construction again requires a globally admissible target density and a mass
function satisfying \(M(v,r)\sim r^3\) at the origin.

\begin{revision}
The sinusoidal density is motivated by condensate models, but the pressure
in equation~\eqref{eq:bec_pressure} is imposed by anisotropic Type-II
balance, not by the Gross--Pitaevskii or isotropic condensate equations.
At $x=\pi$, $\rho=0$ but $P_t=a/2$; the DEC therefore fails near this
boundary, and vanishing density does not establish smooth matching to
vacuum. An admissible exterior or transition layer must replace the
negative-density region, with the appropriate junction conditions.
\end{revision}

\begin{revision}
\section{Cosmological analogue of the reconstruction}
\label{sec:cosmology}
\end{revision}

The interacting reconstruction developed above can also be formulated in a
homogeneous and isotropic setting. This extension is useful because the
continuity equations governing spherical collapse and cosmological evolution
have closely related structures, although the physical interpretation of the
conversion parameter is different. In the collapse problem, the function
\(\beta(v,r)\) is dimensionless because it appears in a radial balance
equation. In cosmology, the corresponding quantity has dimensions of inverse
time. We therefore denote it by \(\Gamma(t)\).

We consider the spatially flat Friedmann--Robertson--Walker metric
\begin{equation}
ds^2=-dt^2+a^2(t)\delta_{ij}dx^i dx^j ,
\end{equation}
where \(a(t)\) is the scale factor and
\begin{equation}
H(t)=\frac{\dot a}{a}
\end{equation}
is the Hubble parameter. For a homogeneous perfect fluid, covariant
conservation gives
\begin{equation}
\dot\rho+3H(\rho+P)=0 .
\label{eq:FRW_continuity}
\end{equation}

\begin{revision}
We now introduce an interacting two-component split consisting of a
first effective component with
\begin{equation}
P_b=w\rho_b,
\end{equation}
and a radiation component satisfying
\begin{equation}
P_r=\frac{1}{3}\rho_r .
\end{equation}
The interacting continuity equations are taken to be
\begin{align}
\dot\rho_r+4H\rho_r
&=
\Gamma(t)\rho_r,
\nonumber\\
\dot\rho_b+3(1+w)H\rho_b
&=
-\Gamma(t)\rho_r .
\label{eq:cosmo_interacting}
\end{align}
Adding the two equations gives the usual conservation equation for the total
source,
\begin{equation}
\dot\rho_{\rm tot}
+
3H(\rho_{\rm tot}+P_{\rm tot})
=
0,
\end{equation}
where
\begin{equation}
\rho_{\rm tot}=\rho_b+\rho_r,
\qquad
P_{\rm tot}=w\rho_b+\frac{1}{3}\rho_r .
\end{equation}
\end{revision}

\begin{revision}
The time-directed source is $\Gamma\rho_r$. A positive $\Gamma$ denotes
transfer into radiation only where $\rho_r>0$; in a signed decomposition its
sign alone is insufficient. The rate is reconstructed rather than derived
from microscopic kinetics. We take $w$ constant and require $w\neq1/3$
for a unique algebraic split.
\end{revision}

\begin{revision}
At fixed background expansion \(H(t)\), the radiation density is formally
\begin{equation}
\rho_r(t)
=
\rho_{0r}
\exp\left[
\int^t
\left(\Gamma(\bar t)-4H(\bar t)\right)d\bar t
\right],
\label{eq:rho_r_cosmo_solution}
\end{equation}
while the first-sector density is
\begin{equation}
\begin{aligned}
\mathcal G(t)&\equiv
\int^t\left[\Gamma(\tilde t)+(3w-1)H(\tilde t)\right]d\tilde t,\\
\rho_b(t)
={}&e^{-\int^t3(1+w)H(\bar t)d\bar t}\\
&\times\left[C-\rho_{0r}\int^t
\Gamma(\bar t)e^{\mathcal G(\bar t)}d\bar t\right],
\end{aligned}
\label{eq:rho_b_cosmo_solution}
\end{equation}
where \(C\) and \(\rho_{0r}\) are integration constants.
\end{revision}

\begin{revision}
We define the target cosmological component by
\begin{equation}
\rho_{\rm new}
=
\rho_b+\rho_r,
\qquad
P_{\rm new}
=
w\rho_b+\frac{1}{3}\rho_r .
\label{eq:cosmo_target_split}
\end{equation}
This target source is not an additional third component; it is the effective fluid produced by the interacting split. Solving
Eq.~\eqref{eq:cosmo_target_split} gives
\begin{equation}
\rho_r
=
\frac{w\rho_{\rm new}-P_{\rm new}}{w-\frac{1}{3}},
\qquad
\rho_b
=
\frac{P_{\rm new}-\frac{1}{3}\rho_{\rm new}}{w-\frac{1}{3}}.
\label{eq:cosmo_algebraic_split}
\end{equation}
\rev{The cosmological split is unique for $w\neq1/3$. At $w=1/3$
it exists only for a radiation-like target $P_{\rm new}=\rho_{\rm new}/3$;
its component densities are then not separately determined.}
\end{revision}

\begin{revision}
Using the radiation equation in Eq.~\eqref{eq:cosmo_interacting}, one finds
\begin{equation}
\Gamma(t)-4H
=
\frac{d}{dt}
\ln\left|w\rho_{\rm new}-P_{\rm new}\right|.
\label{eq:cosmo_log_relation}
\end{equation}
If the target component is separately conserved,
\begin{equation}
\dot\rho_{\rm new}
+
3H(\rho_{\rm new}+P_{\rm new})
=
0,
\label{eq:cosmo_target_conservation}
\end{equation}
then Eq.~\eqref{eq:cosmo_log_relation} gives
\begin{equation}
\Gamma(t)
=
\frac{
wH\rho_{\rm new}
-
(3w+4)H P_{\rm new}
-
\dot P_{\rm new}
}{
w\rho_{\rm new}-P_{\rm new}
}.
\label{eq:Gamma_cosmo}
\end{equation}
This formula is the cosmological analogue of the Husain and
averaged-pressure reconstruction formulae. The correspondence is structural:
\(H(t)\) plays the role of the inverse length scale appearing in the radial
collapse problem, while time derivatives replace radial derivatives. The
dimensionless quantity in cosmology is therefore \(\Gamma/H\), not
\(\Gamma\) itself.
\end{revision}

The cosmological construction should be interpreted as an effective
phenomenological analogue rather than as a direct model of the black-hole
interior. Its purpose is to show that the same inverse-reconstruction logic
can be applied whenever the total conservation law is preserved while the
individual effective components exchange energy.

\begin{revision}
\subsection{Positivity and a Friedmann-consistent comparison}
\label{subsec:cosmo_positivity}
For $\rho>0$, nonnegative component densities require
\begin{equation}
\min(w,1/3)\leq \frac{P}{\rho}\leq\max(w,1/3).
\end{equation}
Positive sectors with $w\geq0$ cannot produce a negative-pressure target.
In particular, $\Gamma/H\to4$ for a de Sitter target is a property of a
signed reconstruction, not a matter--radiation explanation of cosmic
acceleration. If the target is the sole source in flat Einstein-FRW
cosmology, its continuity equation must be supplemented by
$3H^2=\rho$.

For a consistent reference history, take
\begin{align}
H^2(a)&=H_0^2[\Omega_{m0}(a/a_0)^{-3}+\Omega_{\Lambda0}],\\
\rho_m(a)&=3H_0^2\Omega_{m0}(a/a_0)^{-3},\quad
\rho_\Lambda=3H_0^2\Omega_{\Lambda0},\\
\rho&=\rho_m+\rho_\Lambda,\quad P=-\rho_\Lambda,
\quad\Omega_{m0}+\Omega_{\Lambda0}=1.
\label{eq:cosmo_reference}
\end{align}
The values $\Omega_{m0}=0.315$ and $\Omega_{\Lambda0}=0.685$ follow the
base-flat-$\Lambda$CDM reference inferred by Planck
\cite{Planck2018Parameters}. They are not fitted parameters of the
reconstruction. Substitution gives
\begin{equation}
\frac{\Gamma}{H}=
\frac{w\rho_m+4(w+1)\rho_\Lambda}{w\rho_m+(w+1)\rho_\Lambda}.
\label{eq:cosmo_reference_rate}
\end{equation}
Therefore $\Gamma/H\to4$, while
$\Gamma/H_0\to4\sqrt{\Omega_{\Lambda0}}$. A nonnegative alternative for the
same expansion is the usual dust--vacuum split, $\rho_b=\rho_m$,
$\rho_X=\rho_\Lambda$, $P_b=0$, $P_X=-\rho_X$, for which $Q_X=0$.
The expansion history alone does not select a unique exchange law.
Observational constraints on an interacting model require identified
physical sectors and their perturbation equations.
\end{revision}

\section{Numerical checks of the reconstruction}
\label{sec:numerics}
\begin{revision}
All figures are generated by the accompanying notebook and scripts. The
numerical checks compare the mass integral with adaptive quadrature,
verify the radial conservation and reconstruction identities, and locate
horizons and zero-component radii with bracketed root solvers. Curves are
broken at poles instead of joining points across a vanishing denominator.
These calculations test the background formulas, not a closed dynamical
matter theory.

For the phase plots, let $L$ be a reference length and define
$\bar r=r/L$, $\bar\rho=L^2\rho$, $\bar b=L^2b$,
$\bar C=C/L^{2/3}$, and $\bar\eta=\eta/L^{2(\gamma-1)}$.
The constant $D$ in the polytropic $W=M'$ expression is dimensionless.
For the static regular-metric comparisons the asymptotic mass $m$ is used
as the length unit.

\subsection{Bag-model and condensate-inspired targets}
Figure~\ref{fig:quark} uses the averaged-pressure function $\beta_{\rm K}$.
The value $\alpha=1$ is excluded, whereas $\alpha=1/3$ gives the constant
value four when $b>0$. All displayed quark decompositions are signed:
$\rho_X<0$, and at $\alpha=1/3$ specifically $\rho_X=-2b$.
Neither a finite radial function nor its large-radius limit removes the
central divergence of the target density for $C\neq0$.

\begin{figure}[tbp]
\centering
\includegraphics[width=\linewidth]{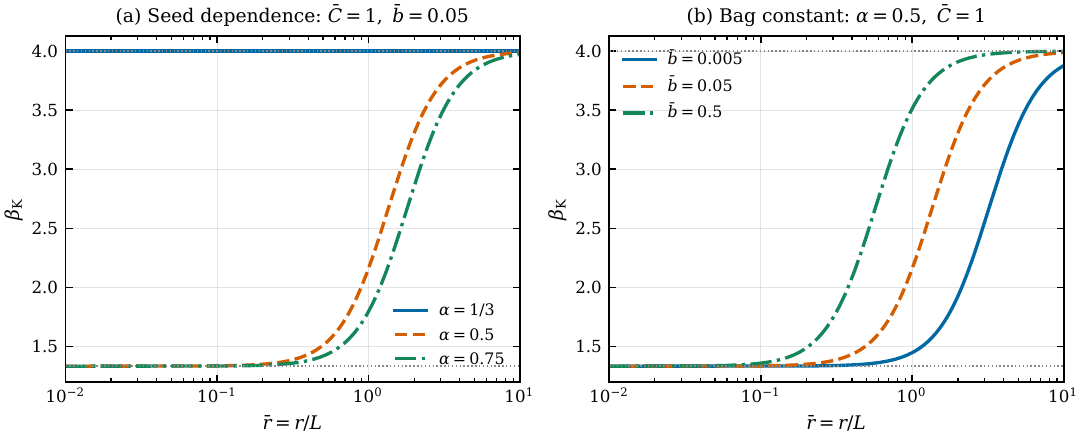}
\caption{Averaged-pressure radial reconstruction for the bag-model target.
(a) $\alpha=1/3,1/2,3/4$ at $(\bar C,\bar b)=(1,0.05)$.
(b) $\bar b=0.005,0.05,0.5$ at $\alpha=1/2$ and $\bar C=1$.
For the nonconstant curves the limiting values are $4/3$ and $4$.
All shown splits have a negative second-component density; they are
algebraic examples rather than positive quark--radiation mixtures.}
\label{fig:quark}
\end{figure}

The condensate-inspired profile in figure~\ref{fig:bec} has nonnegative
central-lobe density for $0\leq x\leq\pi$, but that condition alone does
not ensure the DEC, component positivity, or valid matching. The first
zero-component radii are $x=2.174626,2.288930,2.380644$ for the three
shown coefficients. These lie before the first density node and make
the normalized radial function singular while leaving the total geometry
finite at those radii.

\begin{figure}[tbp]
\centering
\includegraphics[width=\linewidth]{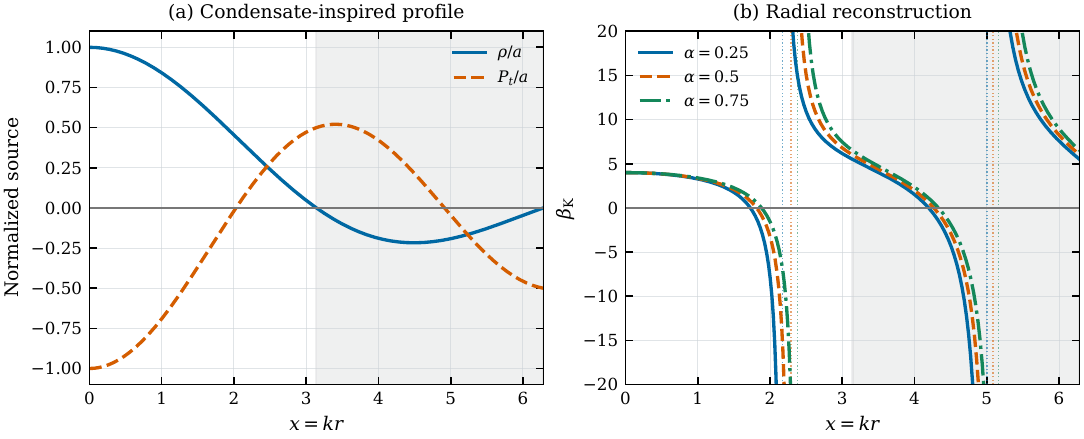}
\caption{Condensate-inspired target over $0\leq x\leq2\pi$.
(a) Density and tangential pressure divided by $a$.
(b) Averaged-pressure radial functions for $\alpha=0.25,0.5,0.75$;
dotted lines locate zeros of the second density.
Shading identifies the negative-total-density continuation. Replacing that
continuation requires an exterior or transition layer and its junction
conditions.}
\label{fig:bec}
\end{figure}

\subsection{Regular-metric source profiles and seed dependence}
We compare the standard static mass functions
\begin{equation}
\begin{aligned}
M_{\rm H}&=\frac{mr^3}{r^3+2m\ell^2},\\
M_{\rm B}&=\frac{mr^3}{(r^2+g^2)^{3/2}},\\
M_{\rm D}&=m\left[1-e^{-(r/r_0)^3}\right],
\end{aligned}
\label{eq:canonical_profiles}
\end{equation}
with $\ell/m=0.8$, $g/m=1.2$, and $r_0/m=1.5$, respectively.
These parameter choices are horizonless members of the Hayward, Bardeen,
and Dymnikova families, used here to compare their regular source profiles.
Their horizon thresholds are $\ell/m=g/m=4/(3\sqrt3)$ and
$(r_0/m)_{\rm crit}=1.3732568\ldots$ for the stated Dymnikova
parametrization. In figure~\ref{fig:RBH}, each density is normalized by its
own central value. All three have $w_t\to-1$ at the center. In the
exterior, $w_t\to2$ for Hayward and $3/2$ for Bardeen, while the Dymnikova
ratio grows despite the exponential decay of its density.

\begin{figure}[tbp]
\centering
\includegraphics[width=\linewidth]{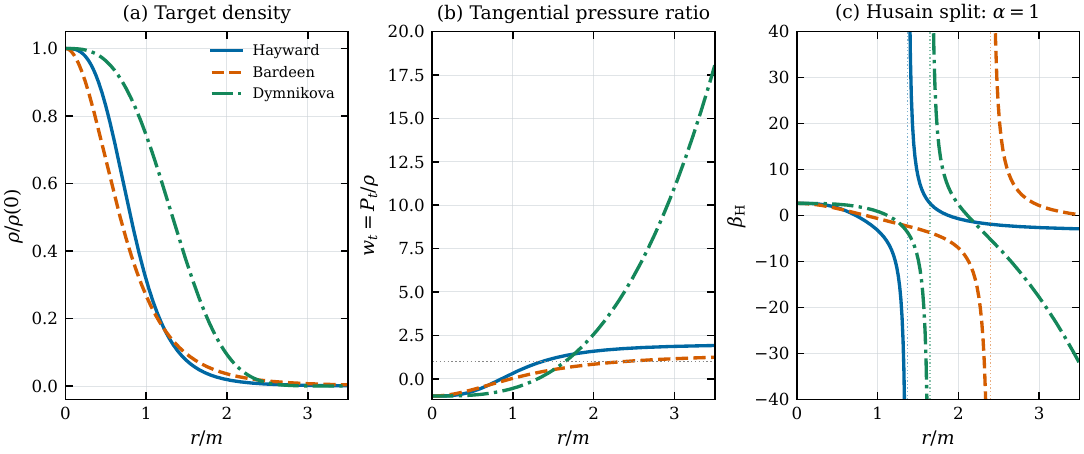}
\caption{Source profiles of the regular-metric families in
equation~\eqref{eq:canonical_profiles} at the stated horizonless parameter
values. (a) Density normalized by its central value. (b) $w_t=P_t/\rho$.
(c) Husain radial reconstruction with $\alpha=1$ and $\chi=1/3$.
Dotted lines mark zero-component poles, not curvature divergences.}
\label{fig:RBH}
\end{figure}

For a fixed nondegenerate split, $\rho_X=0$ when $w_t=\alpha$.
Figure~\ref{fig:alpha} illustrates this dependence for the Hayward profile.
For $\alpha=0,0.25,0.5,0.75,1$, the roots are
$r_\star/m=0.861774,0.970571,1.085767,1.214636,1.367981$.
The source geometry is unchanged by varying $\alpha$; only its partition
and normalized radial function change.

\begin{figure}[tbp]
\centering
\includegraphics[width=\linewidth]{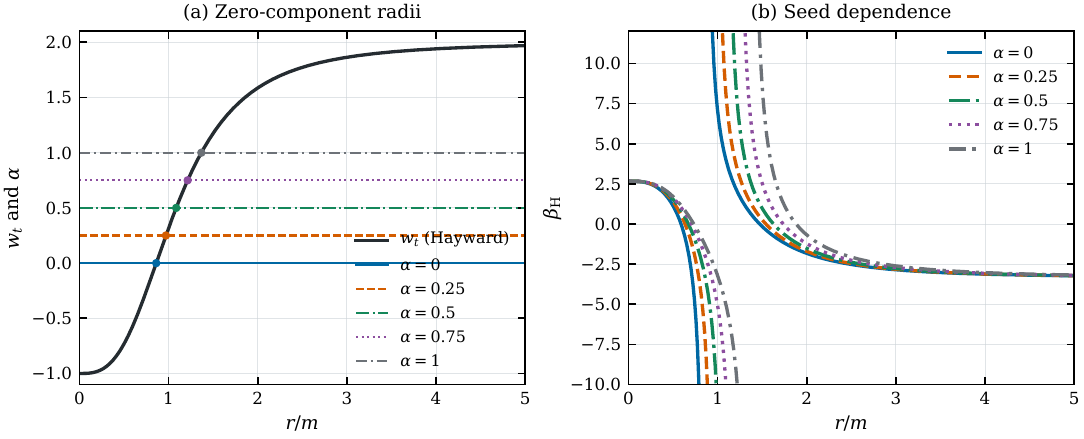}
\caption{Husain seed dependence for the Hayward source with $\ell/m=0.8$.
(a) Intersections of $w_t(r)$ with the displayed $\alpha$ values.
(b) Corresponding radial functions, broken at zero-component poles.
The algebraic degeneracy of this Husain split is instead $\alpha=1/3$,
which is not plotted.}
\label{fig:alpha}
\end{figure}

\subsection{Cosmological comparison}
Both panels of figure~\ref{fig:cosmo} use the same Friedmann-consistent
matter-plus-vacuum history in equation~\eqref{eq:cosmo_reference}.
For $w=0$, $\Gamma/H=4$ at all times; the other curves approach unity
in the matter-dominated limit and four in the vacuum-dominated limit.
The late-time limit in the right panel is instead
$4\sqrt{0.685}=3.310589\ldots$ because it is normalized by $H_0$.
All displayed decompositions are signed. The positive dust--vacuum split
with zero exchange reproduces the same background.

\begin{figure}[tbp]
\centering
\includegraphics[width=\linewidth]{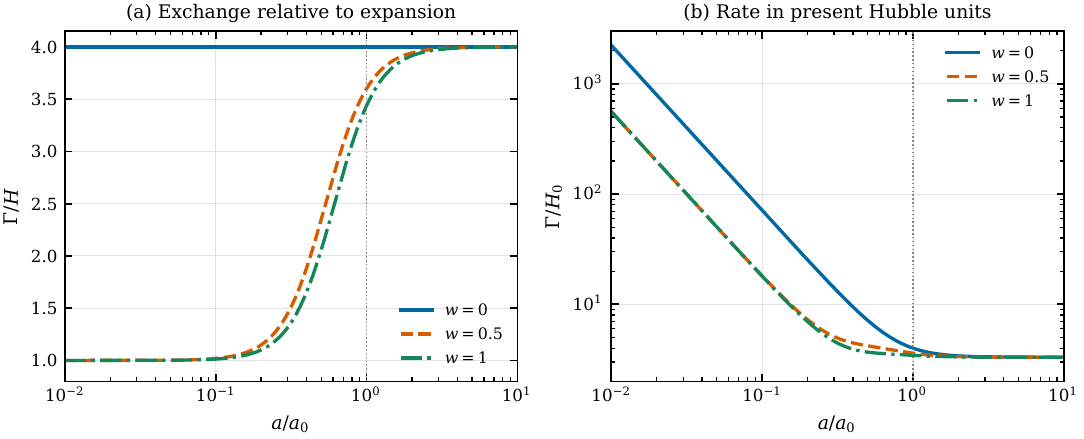}
\caption{Signed cosmological reconstruction for
$\Omega_{m0}=0.315$, $\Omega_{\Lambda0}=0.685$, and $w=0,0.5,1$.
(a) $\Gamma/H$ versus $a/a_0$.
(b) $\Gamma/H_0$ on logarithmic axes.
The limiting constants differ because $H/H_0\to\sqrt{\Omega_{\Lambda0}}$.
These curves are algebraic reconstructions, not a positive
matter--radiation explanation of acceleration.}
\label{fig:cosmo}
\end{figure}

\subsection{Polytropic branches}
Figure~\ref{fig:poly} distinguishes a singular lower branch boundary from
a zero of a reconstructed component. The density branches use
$(\gamma,\bar\eta,D)=(5/3,0.3,1),(2,0.5,1),(5/3,0.8,1)$,
with $\bar r_{\min}=0.573266,1,1.196279$.
For the first branch, the $\alpha=0.25,0.5,0.75$ radial functions have
zero-component poles at $\bar r=1.916829,1.306763,1.082277$,
respectively. These differ from the density singularity at $\bar r_{\min}$.

\begin{figure}[tbp]
\centering
\includegraphics[width=\linewidth]{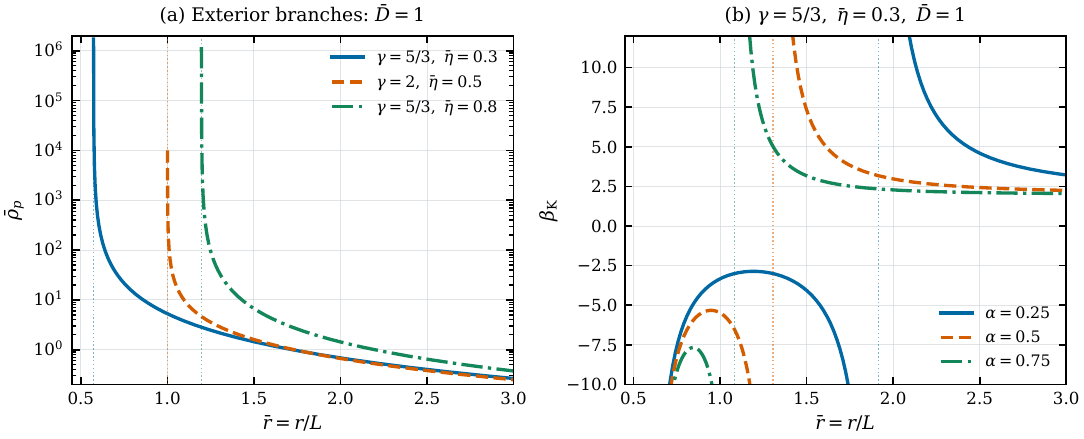}
\caption{Positive real exterior branches of the polytropic target.
(a) Density for the three parameter sets stated in the text; dotted lines
mark their singular lower boundaries.
(b) Averaged-pressure reconstruction for
$(\gamma,\bar\eta,D)=(5/3,0.3,1)$ and $\alpha=0.25,0.5,0.75$.
A finite plotted branch does not constitute a complete regular core.}
\label{fig:poly}
\end{figure}

\subsection{Energy conditions of the comparison profiles}
Figure~\ref{fig:EC} tests the static sources in
equation~\eqref{eq:canonical_profiles}. They have $\rho\geq0$ and
$\rho+P_t\geq0$ at every radius, but the DEC and SEC depend on radius.
In Hayward, Bardeen, and Dymnikova order, the SEC fails below
$r/m=0.861774,0.979796,1.310371$ and the DEC fails above
$r/m=1.367981,2.4,1.650964$. These DEC violations contrast with the
explicit rational-density target, whose total DEC holds everywhere.
An accreting extension must additionally obey $\sigma\geq0$; static
radial plots do not test that condition.

\begin{figure}[tbp]
\centering
\includegraphics[width=\linewidth]{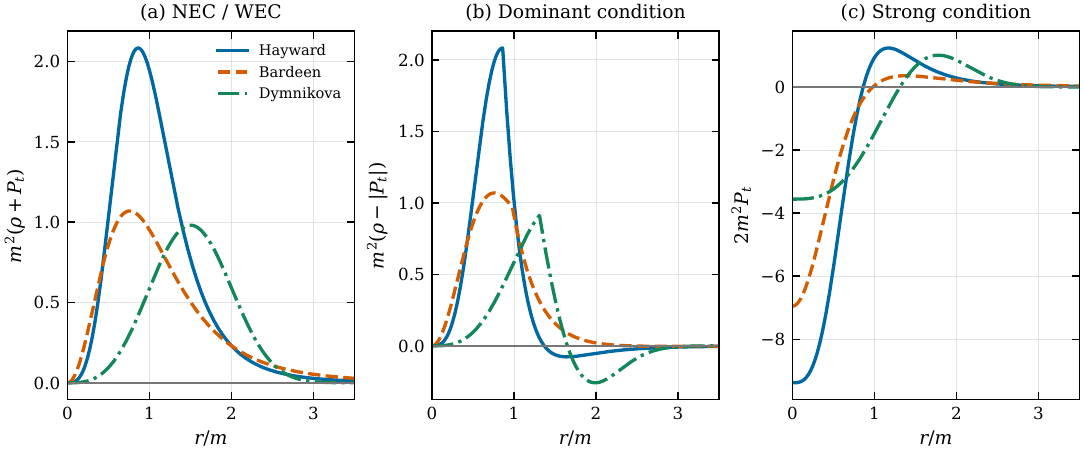}
\caption{Energy-condition combinations for the static sources and
parameters in figure~\ref{fig:RBH}.
(a) $m^2(\rho+P_t)$; (b) $m^2(\rho-|P_t|)$; (c) $2m^2P_t$.
The WEC and NEC hold, the DEC fails beyond the indicated transition
radii, and the SEC fails near the center. SEC violation alone does not
remove the null-convergence and global hypotheses of the Penrose theorem.}
\label{fig:EC}
\end{figure}

\subsection{Hayward trapping-horizon evolution}
As a separate dynamical illustration, choose
\begin{equation}
\begin{aligned}
M(v,r)&=\frac{m(v)r^3}{r^3+2m(v)\ell^2},\\
m(v)&=m_0+\frac{m_\infty-m_0}{2}
\left[1+\tanh\!\left(\frac{v-v_c}{\tau}\right)\right].
\end{aligned}
\label{eq:hayward_smooth_history}
\end{equation}
Here $m_0>0$ and $m_\infty>m_0$, so $\dot M\geq0$.
The positive seed avoids the nonuniform central limit at a zero-mass
startup: for every $m>0$ the center has $\rho(0)=3/\ell^2$,
whereas $m=0$ would be flat. The horizons obey
\begin{equation}
r_h^3-2m(v)r_h^2+2m(v)\ell^2=0.
\end{equation}
Their degenerate point satisfies
\begin{equation}
\begin{aligned}
m_\star&=\frac{3\sqrt3}{4}\ell,\qquad r_c=\sqrt3\ell,\\
v_{\rm form}&=v_c+\tau\operatorname{arctanh}
\left[\frac{2(m_\star-m_0)}{m_\infty-m_0}-1\right].
\end{aligned}
\end{equation}
For the parameters in figure~\ref{fig:hor}, the initial state is
horizonless and two marginal-sphere branches form at
$v_{\rm form}/L=5.320405$. These are trapping horizons, not a computed
event horizon.

\begin{figure}[tbp]
\centering
\includegraphics[width=\linewidth]{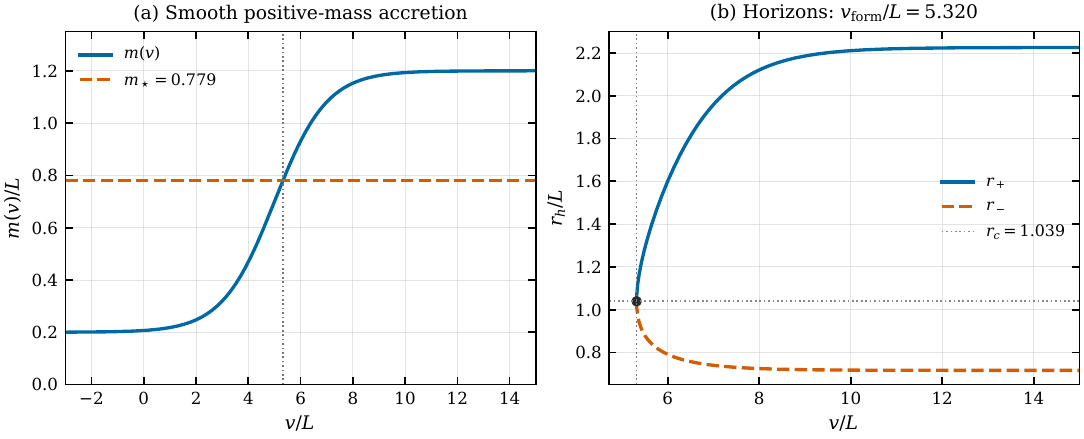}
\caption{Hayward accretion with $\ell/L=0.6$, $m_0/L=0.2$,
$m_\infty/L=1.2$, $v_c/L=5$, and $\tau/L=2$.
(a) Smooth mass history and threshold $m_\star/L=0.779423$.
(b) Inner and outer trapping horizons, meeting at
$r_c/L=1.039230$ and $v_{\rm form}/L=5.320405$.}
\label{fig:hor}
\end{figure}
\end{revision}

\begin{revision}
\section{Observational comparison for the static endpoint}
\label{sec:observations}

Stellar orbits and horizon-scale imaging constrain the exterior geometry of
compact objects. The detection of Schwarzschild precession in the orbit of
S2 supports a compact central mass with a relativistic gravitational field
\cite{GRAVITY2020S2}. The Event Horizon Telescope (EHT) measurements of
Sgr~A$^*$ additionally constrain the size of its shadow after accounting for
the relation between the observed emission ring and the critical null
geodesics \cite{EHTSgrA2022VI}. These measurements do not determine the
stress tensor behind the horizon. They can nevertheless test the explicit
profile in section~\ref{sec:explicit_dynamic}, since its density extends
outside the horizon and modifies the metric near the photon orbit.

Consider its stationary limit, $\rho_c(v)\to\rho_\infty$, with asymptotic
mass $M_\infty$ and lapse
\begin{equation}
\begin{aligned}
F(r)&=1-\lambda\frac{{\cal I}(x)}{x},\\
\lambda&=\rho_\infty r_0^2,\qquad
q\equiv\frac{r_0}{M_\infty}=\frac{4\sqrt{2}}{\pi\lambda}.
\end{aligned}
\label{eq:obs_lapse}
\end{equation}
Here $M_\infty$ is the mass expressed as a gravitational length, and the
comparison varies $q$ at fixed $M_\infty$. The black-hole endpoint exists
for $0<q\le q_c=0.56748639\ldots$, with equality at the degenerate horizon;
$q=0$ denotes the Schwarzschild limit of the exterior geometry. We assume
that the observed photons follow null geodesics of this metric, with
negligible refraction and no additional effective electromagnetic metric.

In the equatorial plane, the conserved photon energy $E$ and angular
momentum $L$ give
\begin{equation}
 \left(\frac{dr}{ds}\right)^2
   =E^2-L^2\frac{F(r)}{r^2},
\end{equation}
where $s$ is an affine parameter. The outer unstable circular orbit obeys
$r_{\rm ph}F'(r_{\rm ph})-2F(r_{\rm ph})=0$, or
\begin{equation}
 \lambda\left[
 \frac{3{\cal I}(x_{\rm ph})}{x_{\rm ph}}
 -\frac{x_{\rm ph}^2}{1+x_{\rm ph}^4}
 \right]=2.
 \label{eq:obs_photon}
\end{equation}
Selecting the root outside the outer horizon with
$d^2(F/r^2)/dr^2<0$, the critical impact parameter and shadow-size
deviation are
\begin{equation}
 b_c=\frac{r_{\rm ph}}{\sqrt{F(r_{\rm ph})}},\qquad
 \delta(q)=\frac{b_c}{3\sqrt{3}M_\infty}-1.
 \label{eq:obs_delta}
\end{equation}
For a distant observer at distance $D$, the angular diameter is
$2b_c/D$ to leading order in $M_\infty/D$.

The origin of the exterior deviation follows directly from
\begin{equation}
 {\cal I}(x)=\frac{\pi}{2\sqrt{2}}-
 \frac{1}{x}+\frac{1}{5x^5}+O(x^{-9}).
\end{equation}
Writing $y=r/M_\infty$, the metric becomes
\begin{equation}
 F=1-\frac{2}{y}
  +\frac{4\sqrt{2}}{\pi}\frac{q}{y^2}
  -\frac{4\sqrt{2}}{5\pi}\frac{q^5}{y^6}
  +O\!\left(\frac{q^9}{y^{10}}\right).
 \label{eq:obs_tail}
\end{equation}
The leading correction has the same radial dependence as the charge term
in the Reissner--Nordstr\"om lapse, although no electric charge has been
introduced here. Expanding equation~\eqref{eq:obs_photon} at small $q$
gives
\begin{equation}
\begin{aligned}
\frac{r_{\rm ph}}{M_\infty}
  &=3-\frac{8\sqrt{2}}{3\pi}q+O(q^2),\\
\delta&=-\frac{2\sqrt{2}}{3\pi}q+O(q^2).
\end{aligned}
\label{eq:obs_smallq}
\end{equation}
Thus the extended density tail decreases the shadow size at fixed
asymptotic mass. The exact result decreases from $\delta=0$ in the
Schwarzschild limit to $\delta=-0.23827$ at the degenerate endpoint.

For a transparent comparison, we use the two fiducial EHT results based
on the \texttt{eht-imaging} reconstruction and its GRMHD calibration,
$\delta=-0.04^{+0.09}_{-0.10}$ with the Keck mass-to-distance prior and
$\delta=-0.08^{+0.09}_{-0.09}$ with the VLTI prior
\cite{EHTSgrA2022VI}. The quoted uncertainties delimit the published
68th-percentile credible intervals. Solving for the values of $q$ whose
predicted $\delta$ lies inside each interval gives
table~\ref{tab:obs_intervals}. The priors are alternative analyses of the
same source; their intervals are not combined as independent measurements.

\begin{table*}[tbp]
\centering
\begin{tabular}{lccc}
\hline
Prior & Published $\delta$ interval
& Overlapping $q$ & Corresponding $\lambda$ \\
\hline
Keck & $[-0.14,\,0.05]$ & $0<q\lesssim0.389$ & $\lambda\gtrsim4.624$ \\
VLTI & $[-0.17,\,0.01]$ & $0<q\lesssim0.452$ & $\lambda\gtrsim3.980$ \\
\hline
\end{tabular}
\caption{Overlap of the static spherical prediction with published EHT
shadow-size intervals. The Schwarzschild limit $q\to0$ is also compatible.
These are interval comparisons, not posterior credible limits on the
parameters of the present model.}
\label{tab:obs_intervals}
\end{table*}

\begin{figure}[tbp]
\centering
\includegraphics[width=\linewidth]{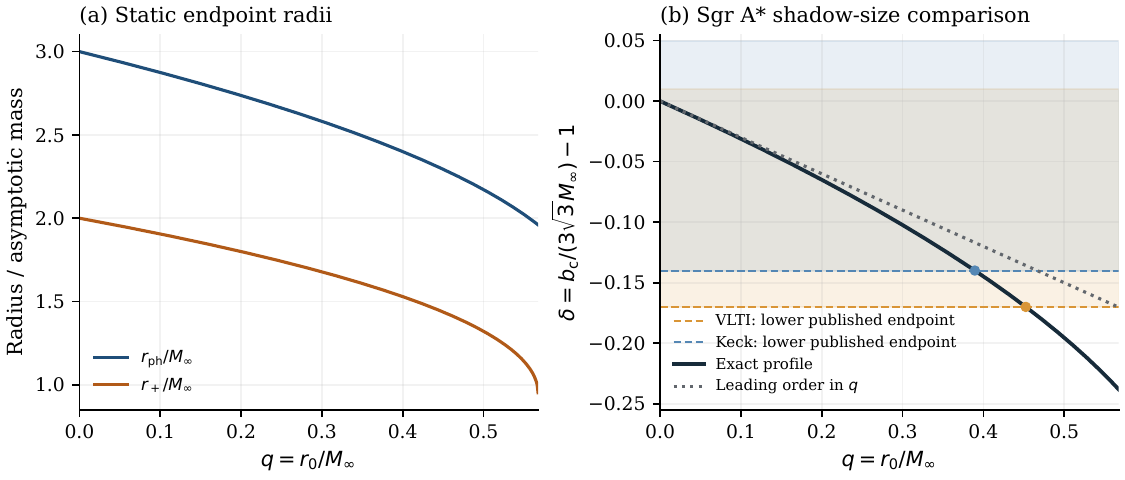}
\caption{Static endpoint geometry and the Sgr~A$^*$ shadow-size comparison.
Panel (a) gives the outer-horizon and unstable-photon-orbit radii in units
of the asymptotic gravitational mass. Panel (b) compares the exact
shadow-size deviation with its leading small-$q$ expansion. Blue and
orange bands show the published Keck and VLTI intervals, respectively;
the marked intersections give the endpoints in
table~\ref{tab:obs_intervals}. The entire horizontal range corresponds to
endpoints with horizons. No EHT likelihood is refitted.}
\label{fig:obs_shadow}
\end{figure}

The endpoint $\lambda=5$ used in the collapse example has
$q=0.36013$ and $\delta=-0.12712$, within both displayed intervals.
Endpoints close to the degenerate horizon give smaller shadows and lie
outside these intervals. A statistical inference for this source would
also require a rotating completion, an emission model appropriate to that
completion, and a joint treatment of the observational and calibration
uncertainties. The present calculation supplies the metric prediction
needed for such an analysis.

The relation between core density and exterior sensitivity is
\begin{equation}
 q=\left(\frac{4\sqrt{2}}
 {\pi\rho_\infty M_\infty^2}\right)^{1/3}.
 \label{eq:obs_density_scale}
\end{equation}
At fixed core density the small-$q$ deviation therefore scales as
$M_\infty^{-2/3}$. Compact high-density cores can produce arbitrarily
small departures from the Schwarzschild shadow within this family.
Moreover, if a different construction has exactly Schwarzschild geometry
throughout the region sampled by the critical rays, its shadow is
independent of changes confined to the interior. The comparison above
tests the assumed exterior density tail; compatibility with observations
is not evidence for matter conversion, central regularity, or a particular
decomposition of the total source.

\end{revision}

\section{Conclusions}
\label{sec:conclusion}
\begin{revision}
We have examined an inverse source construction in generalized Vaidya
spacetimes. A prescribed mass profile fixes the total Type-II source, while
a choice of two tangential equations of state fixes its component densities
and radial reconstruction function. The construction determines a source
partition for a chosen geometry; it does not predict a microscopic matter
transition or the evolution from ordinary baryonic initial data.

Three distinctions control its interpretation. First, the radial function
$\beta$ does not determine a covariant exchange vector without a choice of
component null flux. Second, an algebraically degenerate split must be
distinguished from a zero of one component, at which only the normalized
radial function can diverge. Third, positivity of the total source does not
imply positivity of its component densities. In particular, positive sectors
with nonnegative tangential pressure cannot reproduce a de Sitter center.
A vacuum-like sector removes this algebraic obstruction for the explicit
finite-density target considered here.

For $\rho=\rho_c(v)/(1+x^4)$, we verified the mass function, finite curvature
invariants, and threshold $\rho_c r_0^2=3.172997024\ldots$ for marginal-sphere
formation. Above threshold the inner and outer roots separate with the
expected square-root scaling. Monotonic accretion gives nonnegative null
flux; the total NEC, WEC, and DEC hold, while TCC/SEC violation is confined
to $x<1$. Finite curvature does not establish geodesic completeness, and the
dynamical center has limited differentiability. No claim of perturbative
stability follows from the background pressure--density gradient ratio.

The phase examples impose additional restrictions. The positive polytropic
branch is an exterior-domain solution, the nonconstant MIT-bag profile is
singular at the origin, and the condensate-inspired density requires a
finite-domain treatment. Matching to another region requires junction
conditions beyond the algebraic reconstruction. The cosmological counterpart
has the same positivity limitation: a signed reconstruction of a
$\Lambda$CDM background is not a physical matter--radiation conversion model.

The stationary endpoint also has a nonvacuum exterior that changes its
photon sphere and shadow. The comparison with published Sagittarius A*
shadow-size intervals gives an explicit way to test that exterior under
stated assumptions. It neither detects a regular core nor selects one
source decomposition. A closed matter theory, its perturbation equations,
and a global causal extension are the remaining requirements for turning
this inverse construction into a predictive formation model.
\end{revision}

\backmatter

\bmhead{Acknowledgements}
E. B. acknowledges the support of INFN, \emph{iniziative specifiche MOONLIGHT-2}.
A. {\"O}. acknowledges the contribution of COST Action CA21106---COSMIC WISPers
in the Dark Universe: Theory, Astrophysics and Experiments (CosmicWISPers),
COST Action CA22113---Fundamental Challenges in Theoretical Physics
(THEORY-CHALLENGES), COST Action CA23130---Bridging High and Low Energies in
Search of Quantum Gravity (BridgeQG), and COST Action CA23115---Relativistic
Quantum Information (RQI), funded by COST (European Cooperation in Science and
Technology). A. {\"O}. also thanks TUBITAK and SCOAP3 for their support.

\section*{Statements and Declarations}

\bmhead{Competing interests}
The authors declare that they have no competing interests.

\bmhead{Data availability}
No external datasets were generated or analysed in this study. All data
supporting the findings are contained within the article.

\bibliography{references}


\begin{thebibliography}{61}
\ifx \bisbn   \undefined \def \bisbn  #1{ISBN #1}\fi
\ifx \binits  \undefined \def \binits#1{#1}\fi
\ifx \bauthor  \undefined \def \bauthor#1{#1}\fi
\ifx \batitle  \undefined \def \batitle#1{#1}\fi
\ifx \bjtitle  \undefined \def \bjtitle#1{#1}\fi
\ifx \bvolume  \undefined \def \bvolume#1{\textbf{#1}}\fi
\ifx \byear  \undefined \def \byear#1{#1}\fi
\ifx \bissue  \undefined \def \bissue#1{#1}\fi
\ifx \bfpage  \undefined \def \bfpage#1{#1}\fi
\ifx \blpage  \undefined \def \blpage #1{#1}\fi
\ifx \burl  \undefined \def \burl#1{\textsf{#1}}\fi
\ifx \doiurl  \undefined \def \doiurl#1{\url{https://doi.org/#1}}\fi
\ifx \betal  \undefined \def \betal{\textit{et al.}}\fi
\ifx \binstitute  \undefined \def \binstitute#1{#1}\fi
\ifx \binstitutionaled  \undefined \def \binstitutionaled#1{#1}\fi
\ifx \bctitle  \undefined \def \bctitle#1{#1}\fi
\ifx \beditor  \undefined \def \beditor#1{#1}\fi
\ifx \bpublisher  \undefined \def \bpublisher#1{#1}\fi
\ifx \bbtitle  \undefined \def \bbtitle#1{#1}\fi
\ifx \bedition  \undefined \def \bedition#1{#1}\fi
\ifx \bseriesno  \undefined \def \bseriesno#1{#1}\fi
\ifx \blocation  \undefined \def \blocation#1{#1}\fi
\ifx \bsertitle  \undefined \def \bsertitle#1{#1}\fi
\ifx \bsnm \undefined \def \bsnm#1{#1}\fi
\ifx \bsuffix \undefined \def \bsuffix#1{#1}\fi
\ifx \bparticle \undefined \def \bparticle#1{#1}\fi
\ifx \barticle \undefined \def \barticle#1{#1}\fi
\bibcommenthead
\ifx \bconfdate \undefined \def \bconfdate #1{#1}\fi
\ifx \botherref \undefined \def \botherref #1{#1}\fi
\ifx \url \undefined \def \url#1{\textsf{#1}}\fi
\ifx \bchapter \undefined \def \bchapter#1{#1}\fi
\ifx \bbook \undefined \def \bbook#1{#1}\fi
\ifx \bcomment \undefined \def \bcomment#1{#1}\fi
\ifx \oauthor \undefined \def \oauthor#1{#1}\fi
\ifx \citeauthoryear \undefined \def \citeauthoryear#1{#1}\fi
\ifx \endbibitem  \undefined \def \endbibitem {}\fi
\ifx \bconflocation  \undefined \def \bconflocation#1{#1}\fi
\ifx \arxivurl  \undefined \def \arxivurl#1{\textsf{#1}}\fi
\csname PreBibitemsHook\endcsname

\bibitem[\protect\citeauthoryear{Oppenheimer and
  Snyder}{1939}]{Oppenheimer:1939ue}
\begin{barticle}
\bauthor{\bsnm{Oppenheimer}, \binits{J.R.}},
\bauthor{\bsnm{Snyder}, \binits{H.}}:
\batitle{{On Continued gravitational contraction}}.
\bjtitle{Phys. Rev.}
\bvolume{56},
\bfpage{455}--\blpage{459}
(\byear{1939})
\doiurl{10.1103/PhysRev.56.455}
\end{barticle}
\endbibitem

\bibitem[\protect\citeauthoryear{Penrose}{1965}]{Penrose:1964wq}
\begin{barticle}
\bauthor{\bsnm{Penrose}, \binits{R.}}:
\batitle{{Gravitational collapse and space-time singularities}}.
\bjtitle{Phys. Rev. Lett.}
\bvolume{14},
\bfpage{57}--\blpage{59}
(\byear{1965})
\doiurl{10.1103/PhysRevLett.14.57}
\end{barticle}
\endbibitem

\bibitem[\protect\citeauthoryear{Hawking and Ellis}{1973}]{Hawking:1973LSS}
\begin{bbook}
\bauthor{\bsnm{Hawking}, \binits{S.W.}},
\bauthor{\bsnm{Ellis}, \binits{G.F.R.}}:
\bbtitle{The Large Scale Structure of Space-Time}.
\bpublisher{Cambridge University Press}, \blocation{???}
(\byear{1973})
\end{bbook}
\endbibitem

\bibitem[\protect\citeauthoryear{Bardeen}{1968}]{Bardeen:1968}
\begin{bchapter}
\bauthor{\bsnm{Bardeen}, \binits{J.M.}}:
\bctitle{Nonsingular general relativistic gravitational collapse}.
In: \bbtitle{Proceedings of the International Conference GR5},
\bconflocation{Tbilisi, USSR}
(\byear{1968})
\end{bchapter}
\endbibitem

\bibitem[\protect\citeauthoryear{Dymnikova}{1992}]{Dymnikova:1992ux}
\begin{barticle}
\bauthor{\bsnm{Dymnikova}, \binits{I.}}:
\batitle{{Vacuum nonsingular black hole}}.
\bjtitle{Gen. Rel. Grav.}
\bvolume{24},
\bfpage{235}--\blpage{242}
(\byear{1992})
\doiurl{10.1007/BF00760226}
\end{barticle}
\endbibitem

\bibitem[\protect\citeauthoryear{Hayward}{2006}]{Hayward:2005gi}
\begin{barticle}
\bauthor{\bsnm{Hayward}, \binits{S.A.}}:
\batitle{{Formation and Evaporation of Nonsingular Black Holes}}.
\bjtitle{Phys. Rev. Lett.}
\bvolume{96},
\bfpage{031103}
(\byear{2006})
\doiurl{10.1103/PhysRevLett.96.031103}
{\href{https://arxiv.org/abs/gr-qc/0506126}{{arXiv:gr-qc/0506126}}}
\end{barticle}
\endbibitem

\bibitem[\protect\citeauthoryear{Ansoldi}{2008}]{Ansoldi:2008jw}
\begin{bchapter}
\bauthor{\bsnm{Ansoldi}, \binits{S.}}:
\bctitle{{Spherical black holes with regular center: A Review of existing
  models including a recent realization with Gaussian sources}}.
In: \bbtitle{{Conference on Black Holes and Naked Singularities}}
(\byear{2008})
\end{bchapter}
\endbibitem

\bibitem[\protect\citeauthoryear{Lan et~al.}{2023}]{Lan:2023rbh}
\begin{barticle}
\bauthor{\bsnm{Lan}, \binits{C.}},
\bauthor{\bsnm{Yang}, \binits{H.}},
\bauthor{\bsnm{Guo}, \binits{Y.}},
\bauthor{\bsnm{Miao}, \binits{Y.-G.}}:
\batitle{{Regular Black Holes: A Short Topic Review}}.
\bjtitle{Int. J. Theor. Phys.}
\bvolume{62}(\bissue{9}),
\bfpage{202}
(\byear{2023})
\doiurl{10.1007/s10773-023-05454-1}
{\href{https://arxiv.org/abs/2303.11696}{{arXiv:2303.11696}}}
{[gr-qc]}
\end{barticle}
\endbibitem

\bibitem[\protect\citeauthoryear{Ayon-Beato and
  Garcia}{1998}]{Ayon-Beato:1998hmi}
\begin{barticle}
\bauthor{\bsnm{Ayon-Beato}, \binits{E.}},
\bauthor{\bsnm{Garcia}, \binits{A.}}:
\batitle{{Regular black hole in general relativity coupled to nonlinear
  electrodynamics}}.
\bjtitle{Phys. Rev. Lett.}
\bvolume{80},
\bfpage{5056}--\blpage{5059}
(\byear{1998})
\doiurl{10.1103/PhysRevLett.80.5056}
{\href{https://arxiv.org/abs/gr-qc/9911046}{{arXiv:gr-qc/9911046}}}
\end{barticle}
\endbibitem

\bibitem[\protect\citeauthoryear{Bronnikov}{2000}]{Bronnikov:2000yz}
\begin{barticle}
\bauthor{\bsnm{Bronnikov}, \binits{K.A.}}:
\batitle{{Comment on `Regular black hole in general relativity coupled to
  nonlinear electrodynamics'}}.
\bjtitle{Phys. Rev. Lett.}
\bvolume{85},
\bfpage{4641}
(\byear{2000})
\doiurl{10.1103/PhysRevLett.85.4641}
\end{barticle}
\endbibitem

\bibitem[\protect\citeauthoryear{Bronnikov}{2001}]{Bronnikov:2000vy}
\begin{barticle}
\bauthor{\bsnm{Bronnikov}, \binits{K.A.}}:
\batitle{{Regular magnetic black holes and monopoles from nonlinear
  electrodynamics}}.
\bjtitle{Phys. Rev. D}
\bvolume{63},
\bfpage{044005}
(\byear{2001})
\doiurl{10.1103/PhysRevD.63.044005}
{\href{https://arxiv.org/abs/gr-qc/0006014}{{arXiv:gr-qc/0006014}}}
\end{barticle}
\endbibitem

\bibitem[\protect\citeauthoryear{Frolov et~al.}{1990}]{Frolov:1988vj}
\begin{barticle}
\bauthor{\bsnm{Frolov}, \binits{V.P.}},
\bauthor{\bsnm{Markov}, \binits{M.A.}},
\bauthor{\bsnm{Mukhanov}, \binits{V.F.}}:
\batitle{{Black Holes as Possible Sources of Closed and Semiclosed Worlds}}.
\bjtitle{Phys. Rev. D}
\bvolume{41},
\bfpage{383}
(\byear{1990})
\doiurl{10.1103/PhysRevD.41.383}
\end{barticle}
\endbibitem

\bibitem[\protect\citeauthoryear{Borde}{1997}]{Borde:1996df}
\begin{barticle}
\bauthor{\bsnm{Borde}, \binits{A.}}:
\batitle{{Regular black holes and topology change}}.
\bjtitle{Phys. Rev. D}
\bvolume{55},
\bfpage{7615}--\blpage{7617}
(\byear{1997})
\doiurl{10.1103/PhysRevD.55.7615}
{\href{https://arxiv.org/abs/gr-qc/9612057}{{arXiv:gr-qc/9612057}}}
\end{barticle}
\endbibitem

\bibitem[\protect\citeauthoryear{Vagnozzi et~al.}{2023}]{Vagnozzi:2022moj}
\begin{barticle}
\bauthor{\bsnm{Vagnozzi}, \binits{S.}}, \betal:
\batitle{{Horizon-scale tests of gravity theories and fundamental physics from
  the Event Horizon Telescope image of Sagittarius A}}.
\bjtitle{Class. Quant. Grav.}
\bvolume{40}(\bissue{16}),
\bfpage{165007}
(\byear{2023})
\doiurl{10.1088/1361-6382/acd97b}
{\href{https://arxiv.org/abs/2205.07787}{{arXiv:2205.07787}}}
{[gr-qc]}
\end{barticle}
\endbibitem

\bibitem[\protect\citeauthoryear{Abdujabbarov
  et~al.}{2016}]{Abdujabbarov:2016hnw}
\begin{barticle}
\bauthor{\bsnm{Abdujabbarov}, \binits{A.}},
\bauthor{\bsnm{Amir}, \binits{M.}},
\bauthor{\bsnm{Ahmedov}, \binits{B.}},
\bauthor{\bsnm{Ghosh}, \binits{S.G.}}:
\batitle{{Shadow of rotating regular black holes}}.
\bjtitle{Phys. Rev. D}
\bvolume{93}(\bissue{10}),
\bfpage{104004}
(\byear{2016})
\doiurl{10.1103/PhysRevD.93.104004}
{\href{https://arxiv.org/abs/1604.03809}{{arXiv:1604.03809}}}
{[gr-qc]}
\end{barticle}
\endbibitem

\bibitem[\protect\citeauthoryear{Toshmatov et~al.}{2017}]{Toshmatov:2017zpr}
\begin{barticle}
\bauthor{\bsnm{Toshmatov}, \binits{B.}},
\bauthor{\bsnm{Stuchl{\'\i}k}, \binits{Z.}},
\bauthor{\bsnm{Ahmedov}, \binits{B.}}:
\batitle{{Generic rotating regular black holes in general relativity coupled to
  nonlinear electrodynamics}}.
\bjtitle{Phys. Rev. D}
\bvolume{95}(\bissue{8}),
\bfpage{084037}
(\byear{2017})
\doiurl{10.1103/PhysRevD.95.084037}
{\href{https://arxiv.org/abs/1704.07300}{{arXiv:1704.07300}}}
{[gr-qc]}
\end{barticle}
\endbibitem

\bibitem[\protect\citeauthoryear{Toshmatov et~al.}{2015}]{Toshmatov:2015wga}
\begin{barticle}
\bauthor{\bsnm{Toshmatov}, \binits{B.}},
\bauthor{\bsnm{Abdujabbarov}, \binits{A.}},
\bauthor{\bsnm{Stuchl{\'\i}k}, \binits{Z.}},
\bauthor{\bsnm{Ahmedov}, \binits{B.}}:
\batitle{{Quasinormal modes of test fields around regular black holes}}.
\bjtitle{Phys. Rev. D}
\bvolume{91}(\bissue{8}),
\bfpage{083008}
(\byear{2015})
\doiurl{10.1103/PhysRevD.91.083008}
{\href{https://arxiv.org/abs/1503.05737}{{arXiv:1503.05737}}}
{[gr-qc]}
\end{barticle}
\endbibitem

\bibitem[\protect\citeauthoryear{Capozziello
  et~al.}{2024}]{Capozziello:2024ucm}
\begin{barticle}
\bauthor{\bsnm{Capozziello}, \binits{S.}},
\bauthor{\bsnm{De~Bianchi}, \binits{S.}},
\bauthor{\bsnm{Battista}, \binits{E.}}:
\batitle{{Avoiding singularities in Lorentzian-Euclidean black holes: The role
  of~atemporality}}.
\bjtitle{Phys. Rev. D}
\bvolume{109}(\bissue{10}),
\bfpage{104060}
(\byear{2024})
\doiurl{10.1103/PhysRevD.109.104060}
{\href{https://arxiv.org/abs/2404.17267}{{arXiv:2404.17267}}}
{[gr-qc]}
\end{barticle}
\endbibitem

\bibitem[\protect\citeauthoryear{Wang and Battista}{2026a}]{Wang:2026jvo}
\begin{barticle}
\bauthor{\bsnm{Wang}, \binits{Z.-L.}},
\bauthor{\bsnm{Battista}, \binits{E.}}:
\batitle{{Energy conditions in static, spherically symmetric spacetimes and
  effective geometries}}.
\bjtitle{JCAP}
\bvolume{07},
\bfpage{106}
(\byear{2026})
\doiurl{10.1088/1475-7516/2026/07/106}
{\href{https://arxiv.org/abs/2604.16545}{{arXiv:2604.16545}}}
{[gr-qc]}
\end{barticle}
\endbibitem

\bibitem[\protect\citeauthoryear{Wang and Battista}{2026b}]{Wang:2026sqr}
\begin{botherref}
\oauthor{\bsnm{Wang}, \binits{Z.-L.}},
\oauthor{\bsnm{Battista}, \binits{E.}}:
{Families of regular spacetimes and energy conditions}
(2026)
{\href{https://arxiv.org/abs/2605.03428}{{arXiv:2605.03428}}}
{[gr-qc]}
\end{botherref}
\endbibitem

\bibitem[\protect\citeauthoryear{Bueno et~al.}{2025a}]{Bueno:2024zsx}
\begin{barticle}
\bauthor{\bsnm{Bueno}, \binits{P.}},
\bauthor{\bsnm{Cano}, \binits{P.A.}},
\bauthor{\bsnm{Hennigar}, \binits{R.A.}},
\bauthor{\bsnm{Murcia}, \binits{{\'A}.J.}}:
\batitle{{Regular black holes from thin-shell collapse}}.
\bjtitle{Phys. Rev. D}
\bvolume{111}(\bissue{10}),
\bfpage{104009}
(\byear{2025})
\doiurl{10.1103/PhysRevD.111.104009}
{\href{https://arxiv.org/abs/2412.02740}{{arXiv:2412.02740}}}
{[gr-qc]}
\end{barticle}
\endbibitem

\bibitem[\protect\citeauthoryear{Bueno et~al.}{2025b}]{Bueno:2024eig}
\begin{barticle}
\bauthor{\bsnm{Bueno}, \binits{P.}},
\bauthor{\bsnm{Cano}, \binits{P.A.}},
\bauthor{\bsnm{Hennigar}, \binits{R.A.}},
\bauthor{\bsnm{Murcia}, \binits{{\'A}.J.}}:
\batitle{{Dynamical Formation of Regular Black Holes}}.
\bjtitle{Phys. Rev. Lett.}
\bvolume{134}(\bissue{18}),
\bfpage{181401}
(\byear{2025})
\doiurl{10.1103/PhysRevLett.134.181401}
{\href{https://arxiv.org/abs/2412.02742}{{arXiv:2412.02742}}}
{[gr-qc]}
\end{barticle}
\endbibitem

\bibitem[\protect\citeauthoryear{Bueno et~al.}{2025c}]{Bueno:2025gjg}
\begin{barticle}
\bauthor{\bsnm{Bueno}, \binits{P.}},
\bauthor{\bsnm{Cano}, \binits{P.A.}},
\bauthor{\bsnm{Hennigar}, \binits{R.A.}},
\bauthor{\bsnm{Murcia}, \binits{{\'A}.J.}},
\bauthor{\bsnm{Vicente-Cano}, \binits{A.}}:
\batitle{{Regular black holes from Oppenheimer-Snyder collapse}}.
\bjtitle{Phys. Rev. D}
\bvolume{112}(\bissue{6}),
\bfpage{064039}
(\byear{2025})
\doiurl{10.1103/qrbb-mdvm}
{\href{https://arxiv.org/abs/2505.09680}{{arXiv:2505.09680}}}
{[gr-qc]}
\end{barticle}
\endbibitem

\bibitem[\protect\citeauthoryear{Ovalle}{2025}]{Ovalle:2025pue}
\begin{botherref}
\oauthor{\bsnm{Ovalle}, \binits{J.}}:
{Interior evolution of regular Schwarzschild black holes}
(2025)
\doiurl{10.1016/j.physletb.2026.140912}
{\href{https://arxiv.org/abs/2509.00816}{{arXiv:2509.00816}}}
{[gr-qc]}
\end{botherref}
\endbibitem

\bibitem[\protect\citeauthoryear{Borissova et~al.}{2025a}]{Borissova:2025msp}
\begin{barticle}
\bauthor{\bsnm{Borissova}, \binits{J.}},
\bauthor{\bsnm{Liberati}, \binits{S.}},
\bauthor{\bsnm{Visser}, \binits{M.}}:
\batitle{{Violations of the null convergence condition in kinematical
  transitions between singular and regular black holes, horizonless compact
  objects, and bounces}}.
\bjtitle{Phys. Rev. D}
\bvolume{111}(\bissue{10}),
\bfpage{104054}
(\byear{2025})
\doiurl{10.1103/PhysRevD.111.104054}
{\href{https://arxiv.org/abs/2502.00548}{{arXiv:2502.00548}}}
{[gr-qc]}
\end{barticle}
\endbibitem

\bibitem[\protect\citeauthoryear{Borissova et~al.}{2025b}]{Borissova:2025hmj}
\begin{barticle}
\bauthor{\bsnm{Borissova}, \binits{J.}},
\bauthor{\bsnm{Liberati}, \binits{S.}},
\bauthor{\bsnm{Visser}, \binits{M.}}:
\batitle{{Timelike convergence condition in regular black-hole spacetimes with
  (anti{\textendash})de Sitter core}}.
\bjtitle{Phys. Rev. D}
\bvolume{112}(\bissue{10}),
\bfpage{104072}
(\byear{2025})
\doiurl{10.1103/rrc9-g1sv}
{\href{https://arxiv.org/abs/2509.08590}{{arXiv:2509.08590}}}
{[gr-qc]}
\end{barticle}
\endbibitem

\bibitem[\protect\citeauthoryear{Muniz et~al.}{2026}]{Muniz:2025ugk}
\begin{barticle}
\bauthor{\bsnm{Muniz}, \binits{C.R.}},
\bauthor{\bsnm{Rebou{\c{c}}as}, \binits{J.A.}},
\bauthor{\bsnm{Oliveira}, \binits{L.T.}},
\bauthor{\bsnm{Sampaio}, \binits{F.T.B.}},
\bauthor{\bsnm{Lustosa}, \binits{F.B.}}:
\batitle{{Regularized black hole solution from a new string cloud source}}.
\bjtitle{Phys. Dark Univ.}
\bvolume{52},
\bfpage{102272}
(\byear{2026})
\doiurl{10.1016/j.dark.2026.102272}
{\href{https://arxiv.org/abs/2511.11419}{{arXiv:2511.11419}}}
{[gr-qc]}
\end{barticle}
\endbibitem

\bibitem[\protect\citeauthoryear{He et~al.}{2023}]{He:2023bme}
\begin{barticle}
\bauthor{\bsnm{He}, \binits{X.}},
\bauthor{\bsnm{Zhu}, \binits{S.}},
\bauthor{\bsnm{Yu}, \binits{Y.}},
\bauthor{\bsnm{Karamat}, \binits{A.}},
\bauthor{\bsnm{Babar}, \binits{R.}},
\bauthor{\bsnm{Ali}, \binits{R.}}:
\batitle{{Deflection angle analysis under the influence of non-plasma medium
  and plasma medium for regular black hole with cosmic string}}.
\bjtitle{Int. J. Geom. Meth. Mod. Phys.}
\bvolume{20}(\bissue{12}),
\bfpage{2350205}
(\byear{2023})
\doiurl{10.1142/S0219887823502055}
\end{barticle}
\endbibitem

\bibitem[\protect\citeauthoryear{Balart et~al.}{2025}]{Balart:2024rtj}
\begin{barticle}
\bauthor{\bsnm{Balart}, \binits{L.}},
\bauthor{\bsnm{Panotopoulos}, \binits{G.}},
\bauthor{\bsnm{Rinc{\'o}n}, \binits{{\'A}.}}:
\batitle{{On new regular charged black hole solutions: Limiting Curvature
  Condition, Quasinormal modes and Shadows}}.
\bjtitle{Annals Phys.}
\bvolume{473},
\bfpage{169865}
(\byear{2025})
\doiurl{10.1016/j.aop.2024.169865}
{\href{https://arxiv.org/abs/2412.00550}{{arXiv:2412.00550}}}
{[gr-qc]}
\end{barticle}
\endbibitem

\bibitem[\protect\citeauthoryear{Koch et~al.}{2026}]{Koch:2025gaw}
\begin{barticle}
\bauthor{\bsnm{Koch}, \binits{B.}},
\bauthor{\bsnm{Olmo}, \binits{G.J.}},
\bauthor{\bsnm{Riahinia}, \binits{A.}},
\bauthor{\bsnm{Rinc{\'o}n}, \binits{{\'A}.}},
\bauthor{\bsnm{Rubiera-Garcia}, \binits{D.}}:
\batitle{{Quasi-normal modes and shadows of scale-dependent regular black
  holes}}.
\bjtitle{JCAP}
\bvolume{03},
\bfpage{048}
(\byear{2026})
\doiurl{10.1088/1475-7516/2026/03/048}
{\href{https://arxiv.org/abs/2506.15944}{{arXiv:2506.15944}}}
{[gr-qc]}
\end{barticle}
\endbibitem

\bibitem[\protect\citeauthoryear{Uktamov
  et~al.}{2026}]{UktamjonUktamov:2026dep}
\begin{barticle}
\bauthor{\bsnm{Uktamov}, \binits{U.}},
\bauthor{\bsnm{{\"O}vg{\"u}n}, \binits{A.}},
\bauthor{\bsnm{Pantig}, \binits{R.C.}},
\bauthor{\bsnm{Ahmedov}, \binits{B.}}:
\batitle{{Horizon-brightened acceleration radiation and optical signatures of
  generic regular black holes from nonlinear electrodynamics}}.
\bjtitle{Eur. Phys. J. C}
\bvolume{86}(\bissue{6}),
\bfpage{631}
(\byear{2026})
\doiurl{10.1140/epjc/s10052-026-15841-7}
{\href{https://arxiv.org/abs/2602.15077}{{arXiv:2602.15077}}}
{[gr-qc]}
\end{barticle}
\endbibitem

\bibitem[\protect\citeauthoryear{Balart et~al.}{2023}]{Balart:2023odm}
\begin{barticle}
\bauthor{\bsnm{Balart}, \binits{L.}},
\bauthor{\bsnm{Panotopoulos}, \binits{G.}},
\bauthor{\bsnm{Rinc{\'o}n}, \binits{{\'A}.}}:
\batitle{{Regular Charged Black Holes, Energy Conditions, and Quasinormal
  Modes}}.
\bjtitle{Fortsch. Phys.}
\bvolume{71}(\bissue{12}),
\bfpage{2300075}
(\byear{2023})
\doiurl{10.1002/prop.202300075}
{\href{https://arxiv.org/abs/2309.01910}{{arXiv:2309.01910}}}
{[gr-qc]}
\end{barticle}
\endbibitem

\bibitem[\protect\citeauthoryear{Davlataliev
  et~al.}{2025}]{Davlataliev:2024mjl}
\begin{barticle}
\bauthor{\bsnm{Davlataliev}, \binits{A.}},
\bauthor{\bsnm{Narzilloev}, \binits{B.}},
\bauthor{\bsnm{Hussain}, \binits{I.}},
\bauthor{\bsnm{Abdujabbarov}, \binits{A.}},
\bauthor{\bsnm{Ahmedov}, \binits{B.}}:
\batitle{{Long-lived quasinormal modes and asymptotic tails of regular
  Schwarzschild-like black holes in the presence of a magnetic field}}.
\bjtitle{Phys. Lett. B}
\bvolume{869},
\bfpage{139868}
(\byear{2025})
\doiurl{10.1016/j.physletb.2025.139868}
{\href{https://arxiv.org/abs/2412.09464}{{arXiv:2412.09464}}}
{[gr-qc]}
\end{barticle}
\endbibitem

\bibitem[\protect\citeauthoryear{L{\"u}tf{\"u}o{\u{g}}lu
  et~al.}{2025}]{Lutfuoglu:2025mqa}
\begin{barticle}
\bauthor{\bsnm{L{\"u}tf{\"u}o{\u{g}}lu}, \binits{B.C.}},
\bauthor{\bsnm{Shermatov}, \binits{A.}},
\bauthor{\bsnm{Rayimbaev}, \binits{J.}},
\bauthor{\bsnm{Matyoqubov}, \binits{M.}},
\bauthor{\bsnm{Sirajiddin}, \binits{O.}}:
\batitle{{Gravitational spectra and wave propagation in regular black holes
  supported by a Dehnen Halo}}.
\bjtitle{Eur. Phys. J. C}
\bvolume{85}(\bissue{12}),
\bfpage{1484}
(\byear{2025})
\doiurl{10.1140/epjc/s10052-025-15234-2}
{\href{https://arxiv.org/abs/2511.22366}{{arXiv:2511.22366}}}
{[gr-qc]}
\end{barticle}
\endbibitem

\bibitem[\protect\citeauthoryear{L{\"u}tf{\"u}o{\u{g}}lu}{2026}]{Lutfuoglu:2026boa}
\begin{barticle}
\bauthor{\bsnm{L{\"u}tf{\"u}o{\u{g}}lu}, \binits{B.C.}}:
\batitle{{Scalar, electromagnetic, and Dirac perturbations of regular black
  holes constituting primordial dark matter}}.
\bjtitle{JCAP}
\bvolume{07},
\bfpage{003}
(\byear{2026})
\doiurl{10.1088/1475-7516/2026/07/003}
{\href{https://arxiv.org/abs/2604.24349}{{arXiv:2604.24349}}}
{[gr-qc]}
\end{barticle}
\endbibitem

\bibitem[\protect\citeauthoryear{L{\"u}tf{\"u}o{\u{g}}lu
  et~al.}{2026}]{Lutfuoglu:2026zel}
\begin{barticle}
\bauthor{\bsnm{L{\"u}tf{\"u}o{\u{g}}lu}, \binits{B.C.}},
\bauthor{\bsnm{Rayimbaev}, \binits{J.}},
\bauthor{\bsnm{Murodov}, \binits{S.}},
\bauthor{\bsnm{Abdullaev}, \binits{M.}},
\bauthor{\bsnm{Akhmedov}, \binits{M.}}:
\batitle{{Ringing regularity: Gravitational perturbations and quasinormal modes
  of Einasto-supported black holes}}.
\bjtitle{Annals Phys.}
\bvolume{494},
\bfpage{170664}
(\byear{2026})
\doiurl{10.1016/j.aop.2026.170664}
{\href{https://arxiv.org/abs/2602.20601}{{arXiv:2602.20601}}}
{[gr-qc]}
\end{barticle}
\endbibitem

\bibitem[\protect\citeauthoryear{L{\"u}tf{\"u}o{\u{g}}lu
  et~al.}{2027}]{Lutfuoglu:2026etg}
\begin{barticle}
\bauthor{\bsnm{L{\"u}tf{\"u}o{\u{g}}lu}, \binits{B.C.}},
\bauthor{\bsnm{Abdullaev}, \binits{M.}},
\bauthor{\bsnm{Javlon}, \binits{R.}},
\bauthor{\bsnm{Jumaniyozov}, \binits{S.}},
\bauthor{\bsnm{Karshiboev}, \binits{S.}}:
\batitle{{Gravitational perturbations of a regular T-dulatiy inspired black
  hole: Quasinormal modes, excitation factors, and time-domain evolution}}.
\bjtitle{JHEAp}
\bvolume{55},
\bfpage{100724}
(\byear{2027})
\doiurl{10.1016/j.jheap.2026.100724}
{\href{https://arxiv.org/abs/2607.07715}{{arXiv:2607.07715}}}
{[gr-qc]}
\end{barticle}
\endbibitem

\bibitem[\protect\citeauthoryear{Rahmatov et~al.}{2026}]{Rahmatov:2026qow}
\begin{barticle}
\bauthor{\bsnm{Rahmatov}, \binits{B.}},
\bauthor{\bsnm{Murodov}, \binits{S.}},
\bauthor{\bsnm{Ahmedov}, \binits{B.}}:
\batitle{{Massive scalar perturbations and spectral signatures of the Bardeen
  regular black hole}}.
\bjtitle{Annals Phys.}
\bvolume{493},
\bfpage{170592}
(\byear{2026})
\doiurl{10.1016/j.aop.2026.170592}
\end{barticle}
\endbibitem

\bibitem[\protect\citeauthoryear{Ali et~al.}{2022}]{Ali:2022tdt}
\begin{barticle}
\bauthor{\bsnm{Ali}, \binits{R.}},
\bauthor{\bsnm{Babar}, \binits{R.}},
\bauthor{\bsnm{Sahoo}, \binits{P.K.}}:
\batitle{{Quantum gravity evolution in the Hawking radiation of a rotating
  regular Hayward black hole}}.
\bjtitle{Phys. Dark Univ.}
\bvolume{35},
\bfpage{100948}
(\byear{2022})
\doiurl{10.1016/j.dark.2022.100948}
{\href{https://arxiv.org/abs/2201.02754}{{arXiv:2201.02754}}}
{[gr-qc]}
\end{barticle}
\endbibitem

\bibitem[\protect\citeauthoryear{Ditta et~al.}{2024}]{Ditta:2024jrv}
\begin{barticle}
\bauthor{\bsnm{Ditta}, \binits{A.}},
\bauthor{\bsnm{Xia}, \binits{T.}},
\bauthor{\bsnm{Ali}, \binits{R.}},
\bauthor{\bsnm{Mustafa}, \binits{G.}},
\bauthor{\bsnm{Mustafa}, \binits{G.}},
\bauthor{\bsnm{Mahmood}, \binits{A.}}:
\batitle{{Thermal properties of Simpson{\textendash}Visser Minkowski core
  regular black holes solution in Verlinde{\textquoteright}s emergent
  gravity}}.
\bjtitle{Phys. Dark Univ.}
\bvolume{43},
\bfpage{101418}
(\byear{2024})
\doiurl{10.1016/j.dark.2023.101418}
\end{barticle}
\endbibitem

\bibitem[\protect\citeauthoryear{Rayimbaev et~al.}{2023}]{Rayimbaev:2023vzk}
\begin{barticle}
\bauthor{\bsnm{Rayimbaev}, \binits{J.}},
\bauthor{\bsnm{Abdujabbarov}, \binits{A.}},
\bauthor{\bsnm{Bardiev}, \binits{D.}},
\bauthor{\bsnm{Ahmedov}, \binits{B.}},
\bauthor{\bsnm{Abdullaev}, \binits{M.}}:
\batitle{{Motion of charged and magnetized particles around regular black holes
  immersed in an external magnetic field in modified gravity}}.
\bjtitle{Eur. Phys. J. Plus}
\bvolume{138}(\bissue{4}),
\bfpage{358}
(\byear{2023})
\doiurl{10.1140/epjp/s13360-023-03979-2}
\end{barticle}
\endbibitem

\bibitem[\protect\citeauthoryear{Alloqulov et~al.}{2026}]{Alloqulov:2025bxh}
\begin{barticle}
\bauthor{\bsnm{Alloqulov}, \binits{M.}},
\bauthor{\bsnm{Shaymatov}, \binits{S.}},
\bauthor{\bsnm{Ahmedov}, \binits{B.}},
\bauthor{\bsnm{Zhu}, \binits{T.}}:
\batitle{{Regular black hole{\textquoteright}s impact on the gravitational
  waveforms from periodic orbits}}.
\bjtitle{Eur. Phys. J. C}
\bvolume{86}(\bissue{2}),
\bfpage{117}
(\byear{2026})
\doiurl{10.1140/epjc/s10052-025-15251-1}
{\href{https://arxiv.org/abs/2508.05245}{{arXiv:2508.05245}}}
{[gr-qc]}
\end{barticle}
\endbibitem

\bibitem[\protect\citeauthoryear{Witten}{1984}]{Witten:1984rs}
\begin{barticle}
\bauthor{\bsnm{Witten}, \binits{E.}}:
\batitle{{Cosmic Separation of Phases}}.
\bjtitle{Phys. Rev. D}
\bvolume{30},
\bfpage{272}--\blpage{285}
(\byear{1984})
\doiurl{10.1103/PhysRevD.30.272}
\end{barticle}
\endbibitem

\bibitem[\protect\citeauthoryear{Farhi and Jaffe}{1984}]{Farhi:1984qu}
\begin{barticle}
\bauthor{\bsnm{Farhi}, \binits{E.}},
\bauthor{\bsnm{Jaffe}, \binits{R.L.}}:
\batitle{{Strange Matter}}.
\bjtitle{Phys. Rev. D}
\bvolume{30},
\bfpage{2379}
(\byear{1984})
\doiurl{10.1103/PhysRevD.30.2379}
\end{barticle}
\endbibitem

\bibitem[\protect\citeauthoryear{Chavanis and Harko}{2012}]{Chavanis:2011uv}
\begin{barticle}
\bauthor{\bsnm{Chavanis}, \binits{P.-H.}},
\bauthor{\bsnm{Harko}, \binits{T.}}:
\batitle{{Bose-Einstein Condensate general relativistic stars}}.
\bjtitle{Phys. Rev. D}
\bvolume{86},
\bfpage{064011}
(\byear{2012})
\doiurl{10.1103/PhysRevD.86.064011}
{\href{https://arxiv.org/abs/1108.3986}{{arXiv:1108.3986}}}
{[astro-ph.SR]}
\end{barticle}
\endbibitem

\bibitem[\protect\citeauthoryear{Harko}{2011}]{Harko:2011zt}
\begin{barticle}
\bauthor{\bsnm{Harko}, \binits{T.}}:
\batitle{{Bose-Einstein condensation of dark matter solves the core/cusp
  problem}}.
\bjtitle{JCAP}
\bvolume{05},
\bfpage{022}
(\byear{2011})
\doiurl{10.1088/1475-7516/2011/05/022}
{\href{https://arxiv.org/abs/1105.2996}{{arXiv:1105.2996}}}
{[astro-ph.CO]}
\end{barticle}
\endbibitem

\bibitem[\protect\citeauthoryear{Vertogradov}{2025a}]{Vertogradov:2025dust}
\begin{barticle}
\bauthor{\bsnm{Vertogradov}, \binits{V.}}:
\batitle{{Regular Black Hole from gravitational collapse of dust and
  radiation}}.
\bjtitle{Phys. Dark Univ.}
\bvolume{48},
\bfpage{101881}
(\byear{2025})
\doiurl{10.1016/j.dark.2025.101881}
{\href{https://arxiv.org/abs/2501.13739}{{arXiv:2501.13739}}}
{[gr-qc]}
\end{barticle}
\endbibitem

\bibitem[\protect\citeauthoryear{Vertogradov}{2025b}]{Vertogradov:2025collapse}
\begin{barticle}
\bauthor{\bsnm{Vertogradov}, \binits{V.}}:
\batitle{{Gravitational collapse and formation of regular black holes:
  Dymnikova, Hayward, and beyond}}.
\bjtitle{Eur. Phys. J. C}
\bvolume{85}(\bissue{8}),
\bfpage{839}
(\byear{2025})
\doiurl{10.1140/epjc/s10052-025-14583-2}
{\href{https://arxiv.org/abs/2504.19292}{{arXiv:2504.19292}}}
{[gr-qc]}
\end{barticle}
\endbibitem

\bibitem[\protect\citeauthoryear{Vertogradov and
  {\"O}vg{\"u}n}{2025}]{Vertogradov:2025jxp}
\begin{barticle}
\bauthor{\bsnm{Vertogradov}, \binits{V.}},
\bauthor{\bsnm{{\"O}vg{\"u}n}, \binits{A.}}:
\batitle{{Regular black hole models in the transition from baryonic matter to
  quark matter}}.
\bjtitle{JCAP}
\bvolume{06},
\bfpage{051}
(\byear{2025})
\doiurl{10.1088/1475-7516/2025/06/051}
{\href{https://arxiv.org/abs/2504.07561}{{arXiv:2504.07561}}}
{[gr-qc]}
\end{barticle}
\endbibitem

\bibitem[\protect\citeauthoryear{Vertogradov
  et~al.}{2025}]{Vertogradov:2025snh}
\begin{barticle}
\bauthor{\bsnm{Vertogradov}, \binits{V.}},
\bauthor{\bsnm{{\"O}vg{\"u}n}, \binits{A.}},
\bauthor{\bsnm{Shatov}, \binits{D.}}:
\batitle{{Formation of regular black hole from baryonic matter}}.
\bjtitle{Chin. Phys. C}
\bvolume{49}(\bissue{11}),
\bfpage{115103}
(\byear{2025})
\doiurl{10.1088/1674-1137/ade95c}
{\href{https://arxiv.org/abs/2502.00521}{{arXiv:2502.00521}}}
{[gr-qc]}
\end{barticle}
\endbibitem

\bibitem[\protect\citeauthoryear{Vertogradov and
  {\"O}vg{\"u}n}{2025}]{Vertogradov:2024seh}
\begin{barticle}
\bauthor{\bsnm{Vertogradov}, \binits{V.}},
\bauthor{\bsnm{{\"O}vg{\"u}n}, \binits{A.}}:
\batitle{{Exact regular black hole solutions with de Sitter cores and Hagedorn
  fluid}}.
\bjtitle{Class. Quant. Grav.}
\bvolume{42}(\bissue{2}),
\bfpage{025024}
(\byear{2025})
\doiurl{10.1088/1361-6382/ada082}
{\href{https://arxiv.org/abs/2408.02699}{{arXiv:2408.02699}}}
{[gr-qc]}
\end{barticle}
\endbibitem

\bibitem[\protect\citeauthoryear{Lambiase et~al.}{2026}]{Lambiase:2026hza}
\begin{botherref}
\oauthor{\bsnm{Lambiase}, \binits{G.}},
\oauthor{\bsnm{{\"O}vg{\"u}n}, \binits{A.}},
\oauthor{\bsnm{Vertogradov}, \binits{V.}}:
{Can finite-density QCD matter support a regular black hole core?}
(2026)
{\href{https://arxiv.org/abs/2605.27170}{{arXiv:2605.27170}}}
{[astro-ph.HE]}.
Accepted for publication in JCAP
\end{botherref}
\endbibitem

\bibitem[\protect\citeauthoryear{Husain}{1996}]{Husain:1995}
\begin{barticle}
\bauthor{\bsnm{Husain}, \binits{V.}}:
\batitle{{Exact solutions for null fluid collapse}}.
\bjtitle{Phys. Rev. D}
\bvolume{53},
\bfpage{1759}--\blpage{1762}
(\byear{1996})
\doiurl{10.1103/PhysRevD.53.R1759}
{\href{https://arxiv.org/abs/gr-qc/9511011}{{arXiv:gr-qc/9511011}}}
\end{barticle}
\endbibitem

\bibitem[\protect\citeauthoryear{Wang and Wu}{1999}]{WangWu1999}
\begin{barticle}
\bauthor{\bsnm{Wang}, \binits{A.}},
\bauthor{\bsnm{Wu}, \binits{Y.}}:
\batitle{{Generalized Vaidya Solutions}}.
\bjtitle{Gen. Rel. Grav.}
\bvolume{31},
\bfpage{107}--\blpage{114}
(\byear{1999})
\doiurl{10.1023/A:1018819521971}
{\href{https://arxiv.org/abs/gr-qc/9803038}{{arXiv:gr-qc/9803038}}}
\end{barticle}
\endbibitem

\bibitem[\protect\citeauthoryear{Vertogradov}{2020}]{Vertogradov:2016gc}
\begin{barticle}
\bauthor{\bsnm{Vertogradov}, \binits{V.}}:
\batitle{{The diagonalization of generalized Vaidya spacetime}}.
\bjtitle{Int. J. Mod. Phys. A}
\bvolume{35}(\bissue{02n03}),
\bfpage{2040033}
(\byear{2020})
\doiurl{10.1142/S0217751X20400333}
\end{barticle}
\endbibitem

\bibitem[\protect\citeauthoryear{Akiyama et~al.}{2022}]{EHTSgrA2022VI}
\begin{barticle}
\bauthor{\bsnm{Akiyama}, \binits{K.}}, \betal:
\batitle{{First Sagittarius A* Event Horizon Telescope Results. VI. Testing the
  Black Hole Metric}}.
\bjtitle{Astrophys. J. Lett.}
\bvolume{930}(\bissue{2}),
\bfpage{17}
(\byear{2022})
\doiurl{10.3847/2041-8213/ac6756}
{\href{https://arxiv.org/abs/2311.09484}{{arXiv:2311.09484}}}
{[astro-ph.HE]}
\end{barticle}
\endbibitem

\bibitem[\protect\citeauthoryear{Kiselev}{2003}]{Kiselev:2002bh}
\begin{barticle}
\bauthor{\bsnm{Kiselev}, \binits{V.V.}}:
\batitle{{Quintessence and black holes}}.
\bjtitle{Class. Quant. Grav.}
\bvolume{20},
\bfpage{1187}--\blpage{1198}
(\byear{2003})
\doiurl{10.1088/0264-9381/20/6/310}
{\href{https://arxiv.org/abs/gr-qc/0210040}{{arXiv:gr-qc/0210040}}}
\end{barticle}
\endbibitem

\bibitem[\protect\citeauthoryear{Visser}{2019}]{Visser2020Kiselev}
\begin{botherref}
\oauthor{\bsnm{Visser}, \binits{M.}}:
{The Kiselev black hole is neither perfect fluid, nor is it quintessence}
(2019)
\doiurl{10.1088/1361-6382/ab60b8}
{\href{https://arxiv.org/abs/1908.11058}{{arXiv:1908.11058}}}
{[gr-qc]}
\end{botherref}
\endbibitem

\bibitem[\protect\citeauthoryear{Vertogradov}{2024}]{Vertogradov:2024grg}
\begin{barticle}
\bauthor{\bsnm{Vertogradov}, \binits{V.}}:
\batitle{{The generalized Vaidya spacetime with polytropic equation of state}}.
\bjtitle{Gen. Rel. Grav.}
\bvolume{56}(\bissue{5}),
\bfpage{59}
(\byear{2024})
\doiurl{10.1007/s10714-024-03244-6}
{\href{https://arxiv.org/abs/2311.15671}{{arXiv:2311.15671}}}
{[gr-qc]}
\end{barticle}
\endbibitem

\bibitem[\protect\citeauthoryear{Aghanim et~al.}{2020}]{Planck2018Parameters}
\begin{barticle}
\bauthor{\bsnm{Aghanim}, \binits{N.}}, \betal:
\batitle{{Planck 2018 results. VI. Cosmological parameters}}.
\bjtitle{Astron. Astrophys.}
\bvolume{641},
\bfpage{6}
(\byear{2020})
\doiurl{10.1051/0004-6361/201833910}
{\href{https://arxiv.org/abs/1807.06209}{{arXiv:1807.06209}}}
{[astro-ph.CO]}
\end{barticle}
\endbibitem

\bibitem[\protect\citeauthoryear{Abuter et~al.}{2020}]{GRAVITY2020S2}
\begin{barticle}
\bauthor{\bsnm{Abuter}, \binits{R.}}, \betal:
\batitle{{Detection of the Schwarzschild precession in the orbit of the star S2
  near the Galactic centre massive black hole}}.
\bjtitle{Astron. Astrophys.}
\bvolume{636},
\bfpage{5}
(\byear{2020})
\doiurl{10.1051/0004-6361/202037813}
{\href{https://arxiv.org/abs/2004.07187}{{arXiv:2004.07187}}}
{[astro-ph.GA]}
\end{barticle}
\endbibitem

\end{thebibliography}

\end{document}